\documentclass[12]{article}
\usepackage{amsfonts, amsmath, amssymb, appendix, natbib, xcolor, graphicx, booktabs}

\graphicspath{{./Figures/}}

\newcommand\beq[1]{ \begin{equation}\label{#1} }
\newcommand{\eeq}{ \end{equation} }
\newcommand{\beqno}{ \[ }
\newcommand{\eeqno}{ \] }
\newcommand\beqa[1]{ \begin{eqnarray} \label{#1} }
\newcommand{\eeqa}{ \end{eqnarray} }
\newcommand{\beqano}{ \begin{eqnarray*} }
\newcommand{\eeqano}{ \end{eqnarray*} }
\newcommand\equ[1]{{\rm (\ref{#1})}}

\title{Tadpole type motion of charged dust in the Lagrange problem with planet Jupiter}

\author{
	Christoph Lhotka$^{1,2}$, Lei Zhou$^{2,3}$ \\
	${}^1$Department of Mathematics,
	University of Rome Tor Vergata, \\
	Via della Ricerca Scientifica 1, 00133 Roma, \underline{Italy},\\
	\texttt{lhotka@mat.uniroma2.it} \\ 
	${}^2$Institute of Astrophysics,
        University of Vienna, \\
        T\"urkenschanzstrasse 17, A-1180 Wien, \underline{Austria}, \\
	\texttt{christoph.lhotka@univie.ac.at} \\
	${}^3$School of Astronomy and Space Science,
	Nanjing University, \\
	163 Xianlin Avenues, 210046 Nanjing, \underline{PR China},\\
	\texttt{leizhou@smail.nju.edu.cn}
}
\begin{document}
\maketitle

\begin{abstract}
We investigate the dynamics of charged dust interacting with the interplanetary 
magnetic field in a Parker spiral type model and subject to the solar wind
and Poynting-Robertson effect in the vicinity of the 1:1 mean motion resonance
with planet Jupiter. We estimate the shifts of the location of the minimum libration
amplitude solutions close to the location of the $L_4$ and $L_5$ points of the 
classical - gravitational - problem and provide the extension of the 'librational
regimes of motion' and the width of the resonance in dependency of the nongravitational
parameters related to the dust grain size and surface potential of the particles.
Our study is based on numerical simulations in the framework of the spatial, 
elliptic restricted three-body problem and semi-analytical estimates obtained by
averaging of Gauss' planetary equations of motion.

\end{abstract}


\section{Introduction}
\label{s:int}

Interplanetary dust particles (IDPs) originate largely from cometary activity
and collisions of asteroids in the solar system and form the interplanetary
background dust cloud. A better understanding of the dynamics of dust
distributions in space will not only help to better understand the current
state of our solar system
\citep[][]{2020JGRA..12528463P, 2021A&A...645A..63Z, 2019ApJ...873L..16P}, 
but will also help to design proper spacecraft
shielding.
 Since impacts of dust on spacecraft may strongly influence the quality
of spacecraft data (e.g. plasma measurements, see \citet{2020PhPl...27j3704L})
a better understanding of the distribution and dynamics of IDPs in the vicinity of 
Jupiter is of particular interest for future space missions (e.g. JUICE and LUCY).
In these outer regions, these tiny objects can originate from
several sources, including Edgeworth-Kuiper Belt objects, Halley-type comets,
Jupiter-family comets, and Oort Cloud comets \citep[][]{2019SSRv..215...34K, 
2016Icar..264..369P}, or
the Centaurs \citep[][]{2019MNRAS.490.2421P}. Additionally, interstellar dust
(ISD) has been detected \citep[][]{2019A&A...626A..37K, 1993Natur.362..428G},
meaning that some particles from the Local Interstellar Cloud can penetrate the
heliosphere and enter the solar system.  
Research on the dynamics of dust in the solar system in recent years also include
the study of the evolution of orbits about comets with arbitrary comae
\citep[][]{2020CeMDA.132...37M}, the development of analytical models
to investigate the dynamics in phase space \citep[][]{2019CeMDA.131...43A}, and
a theoretical study on the dissipative Kepler problem with a family of 
singular drags \citep[][]{2020CeMDA.132...17M}. In
\citep[][]{2019CNSNS..76...71F} the authors study the secular dynamics
around small bodies with solar radiation pressure with application
to asteroids, the effect of resonances in the Earth space ennivronment
has been investigated in \citep[][]{2020CNSNS..8405185C}.

In the present work we focus on the dynamcis of charged dust
within co-orbital configuration with a planet. In
\citet{2018AaA...609A..57L,2018AaA...614A..97L} the authors investigate the
orbital evolution of dust released from co-orbital asteroids with planet
Jupiter that form an arc, mainly composed of grains in the size range 4-10
microns. The arc is distributed more widely in the azimuthal direction, with
two peaks that are azimuthally displaced from the equilibrium position of the
pure gravitational problem. The study strongly focuses on the leading Lagrange
point $L_4$ stating that the dust distribution in the vicinity of the region
$L_5$ is similar. However, in \citet{2015Icar..250..249L} we already found an
asymmetry between the Lagrange points $L_4$ and $L_5$ that is caused by the
non-gravitational forces, i.e. by the so-called Poynting-Robertson effect, and
in \citet{2021A&A...645A..63Z}, the effect of asymmetry between the leading and
trailing equilibria has also been found in case of planet Venus. However, while
our study in \citet{2021A&A...645A..63Z} already included the effect of charge,
our study of micron sized dust in the orbit of Jupiter
\citep{2015Icar..250..249L} did not include the effect of the interplanetary
magnetic field, like it has already been done in \citet{2018AaA...609A..57L}.
Lorentz force due to the interaction with the magnetic fields cannot be
neglected if the charge-to-mass ratios become large (which is true for micron
sized particles). The photoelectric effect and charging currents due to space
plasmas result in (mostly) positively charged dust grains, with surface
potential about $5$ Volts \citep[][]{2014PhR...536....1M}. For a detailed
analysis of different charging mechanisms see also \citet{2020PhPl...27j3704L}.
We notice that the secular effect on charged particles out of resonance has
already been investigated in \citet{2016ApJ...828...10L}, where the authors
identify the normal component of the field to trigger secular drift in
semi-major axes.  In addition, a Parker spiral model of the field results in a
normal component in the Lorentz force that will strongly affect the inclination
of the orbital planes. This is also true for charged particles in mean motion
resonance with planet Jupiter. As it has been shown in
\citet{2019CeMDA.131...49L} the interaction with the interplanetary magnetic
field destabilizes the orbits of charged dust grains also in outer mean motion
resonance with the planet.  \\

In the present study we aim to complement these previous results. We study the 
dynamics of micron sized dust and co-orbital with planet Jupiter, and i) 
include the role of charge together with the effect of the interplanetary 
magnetic field and ii) also perform the analysis of dust grain motion close to 
the  trailing Lagrange point $L_5$. In addition, we provide a thorough 
analysis of the extent of the resonant regime of motion in dependency of the 
system parameters, i.e. charge-to-mass ratio and size-to-mass ratio, and 
provide information about typical times of temporary capture in tadpole and
also co-orbital type of motions. \\

This work is organized as follows. In Section~\ref{s:mod} we state the notation
and dynamical problem that we are going to use in our study. The analysis of
the numerical simulation data is given in Section~\ref{s:stud}. The discussion
of certain aspects of the dynamics based on Gauss' equations of motion is
provided in Section~\ref{s:ana}. The summary of the results and conclusions can
be found in Section~\ref{s:sum}.

\section{Notation and set-up}
\label{s:mod}

\begin{table}[]
\centering
\begin{tabular}{l | l | l}
\toprule
symbol & values & reference \\
\hline
	$a_1$ & $ a_{J}=5.205AU$  & (a) \\
$B_0$ & 3nT & \cite{2012bsw..book.....M} \\
$\eta$ & $1/3$  & \cite{2014MNRAS.443..213K} \\
$i_0$ & $7.15^o$ & \cite{2005ApJ...621L.153B} \\
$m_0$ & $M_\odot$ & (a)  \\
	$m_1$ & $m_J=0.001M_\odot$ & (a) \\
$r_0$ & $1AU$ & \cite{2012bsw..book.....M} \\
$\Omega_0$ & $73.5^o$ & \cite{2005ApJ...621L.153B} \\
$\Omega_s^{-1}$ & $24.47d$ & \cite{2012bsw..book.....M} \\
$Q$ & $1$ & \cite{1994Icar..110..239B} \\
$\rho$ & $2.8g/cm^3$ & \cite{1994Icar..110..239B} \\
$u_{sw}$ &400km/s& \cite{2012bsw..book.....M}  \\
\hline
\end{tabular}
	\caption{Parameters, (a) taken from
	{https://nssdc.gsfc.nasa.gov/planetary/factsheet/}.}
\label{t:pars}
\end{table}

The orbital evolution of a charged dust grain is determined by the equation 
of motion

\beq{e:eom}
\ddot{\vec r} + \mu \frac{\vec r}{r^3}=\vec F \ .
\eeq

\noindent Here, $\vec r$ is the position of the dust grain in a heliocentric
coordinate frame with $r=|\vec r|$, and $\mu=G m_0$ with gravitational constant
$G$ and mass of the sun $m_0$.  In absence of perturbing force $\vec F$ (per
unit mass) the particle is assumed to move on a Kepler orbit with constant
orbital elements $a$ (semi-major axis), $e$ (eccentricity), $i$ (inclination),
$\omega$ (perihelion argument), $\Omega$ (ascending node longitude),  mean
anomaly $M$, and mean motion $n$. 
Let $\vec F$ be decomposed into $\vec F=\vec F_0+...+\vec F_3$ with:

\beqa{e:F}
\vec F_0 &=& -\beta\mu\frac{\vec r}{\vec r}{r^3} \nonumber \\
\vec F_1 &=& -\nabla\mu_1\left(\frac{\vec r_1.\vec r}{r_1^3}-\frac{1}{\Delta} -\frac{1}{r}\right) \nonumber \\
\vec F_2 &=& -\frac{\mu\beta}{r^2}\left(1+\frac{\eta}{Q}\right)
\left(
\frac{(\dot{\vec r}\cdot\vec g_r)\vec g_r+\dot{\vec r}}{c}
\right) \nonumber , \nonumber \\
\vec F_3 &=& \gamma\left(\dot{\vec r}-\vec u_{sw}\right)\times\vec B \ .
\eeqa

\noindent Here, $\vec F_0$ is due to solar radiation pressure, $\vec F_1$
is the gravitational force from planet Jupiter, $\vec F_2$ is stemming from the
so-called Poynting-Robertson effect and solar wind drag
\citep[][]{2012MNRAS.421..943K, 2014MNRAS.443..213K}, and $\vec F_3$ is the
Lorentz force term stemming from the interaction of the charged particle with
the interplanetary magnetic field \citep[][]{2019AnGeo..37..299L}. We denote
by $m=4\pi/3\rho R^3$ the mass of a spherical dust grain of radius $R$ with
density $\rho$, and by $q=4\pi\varepsilon_0UR$ its charge for given surface
potential charge $U$ and dielectric constant $\varepsilon_0$. The equations
\equ{e:F} enter the remaining quantities  (in the order of appearance): the
parameter $\beta\propto R^{-1}$,  which is the ratio between the magnitudes of
solar radiation pressure and gravitational force due to the sun:

\beqno
 \beta = \frac{S Q \pi R^2}{c}\left(\frac{r_0}{r}\right)^{-2} / \frac{\mu m}{r^2}
\eeqno

(we notice that both forces are proportional to the inverse square of the
distance from the sun). Setting the solar flux constant
$S=1360.8kgs^{-3}$ at reference distance $r_0=1AU$, the dimensionless
efficiency factor $Q=1$, speed of light $c=299792458m/s$,
$\mu=1.327\times10^{20}m^3s^{-2}$, and $m=11.7286gcm^{-3}$, we find
$\beta=0.205/R$, with $R$ given in micro-meters. The additional quantities are
the gravitational mass of Jupiter $\mu_1=Gm_1$, the heliocentric position
vector for planet Jupiter, $\vec r_1$, the distance between Jupiter and the
dust grain $\Delta=\|\vec r - \vec r_1\|$, the solar wind {efficiency} factor $\eta$, 
$\vec g_r=\vec r/r$,  the
charge-to-mass ratio $\gamma\propto U R^{-2}$, the solar wind speed $\vec
u_{sw}=u_{sw}\vec g_r$, and the interplanetary magnetic field vector $\vec B$.
Our main focus lies in the role of parameters $\beta$, $\gamma$ related to the
physical parameters size and charge. Assuming spherical particles, and  using the
actual values for the various parameters entering \equ{e:F} that are given in
Tab.~\ref{t:pars}, we find:

\beq{e:betgam}
\beta = 0.205/R \quad \gamma = 0.0094U/R^2 \ ,
\eeq

\noindent with $R$ given in micro-meters and $U$ given in Volts. We are left to
specify the form of the interplanetary magnetic field vector:

\beq{e:Bfldbis}
\vec B = \frac{B_0 r_0^2}{r^2}
\left(\frac{\vec r}{r} - \frac{\Omega_s} {u_{sw}} {\vec g}_{\tilde{z}} \times {\vec r} \right) \tanh\Bigl(\alpha \frac{{\vec r} \cdot {\vec g}_{\tilde z}}{r} \Bigr) \ ,
\eeq

\noindent with background magnetic field strength $B_0$ defined at reference
distance $r_0$, and solar rotation rate $\Omega_s$. The form of the $\vec B$
field enters the parameter $\alpha$ to model the sign change when crossing the
equatorial plane of the sun\footnote{We assume that the magnetic dipole axis
and the rotation axis of the sun are aligned and that the equatorial plane of
the sun therefore coincides with the zero current sheet.}, and the unit vector
along magnetic north ${\vec g}_{\tilde z}$ that is related to the heliocentric
reference frame $\{\vec g_x,\vec g_y ,\vec g_z\}$ by

\beqno
\vec g_{\tilde z} =
\sin\left(i_0\right)\left[
\sin\left(\Omega_0\right)\vec g_{x} -
\cos\left(\Omega_0\right)\vec g_{y}\right] +
\cos\left(i_0\right) \vec g_{z} \ .
\eeqno

\begin{figure} \centering
\resizebox{1.0\hsize}{!}{\includegraphics[width=0.75\linewidth]{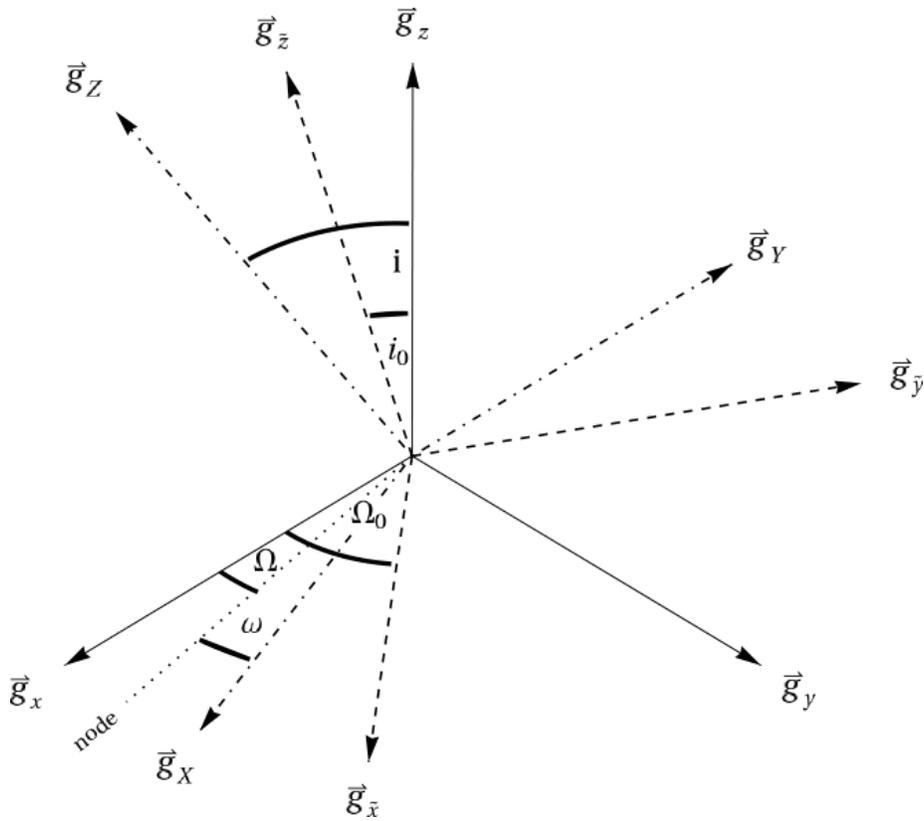}}
	\caption{Geometry of the problem: ecliptic frame $(x,y,z)$,
	equatorial frame $(\tilde x, \tilde y, \tilde z)$, orbital frame
	$(X, Y, Z)$. The nodal line refers to the intersection between the ecliptic and
	orbital planes. For the definitions of the angles, see text.}
\label{f:geo}
\end{figure}

\noindent Here, $i_0$ denotes the inclination between the rotation axis of the
sun and the ecliptic pole and $\Omega_0$ is the angle between the direction of
the vernal equinox and the line of nodes between the equatorial and ecliptic
planes, respectively. Please see Fig.~\ref{f:geo} for visualization of the
various reference frames. The actual parameters used in our study are
summarized in Tab.~\ref{t:pars}, for further details on the model, see
\citet{2019CeMDA.131...49L}.

\section{Numerical study and results}
\label{s:stud}

\begin{figure} \centering
\resizebox{1.0\hsize}{!}{\includegraphics[width=\linewidth]{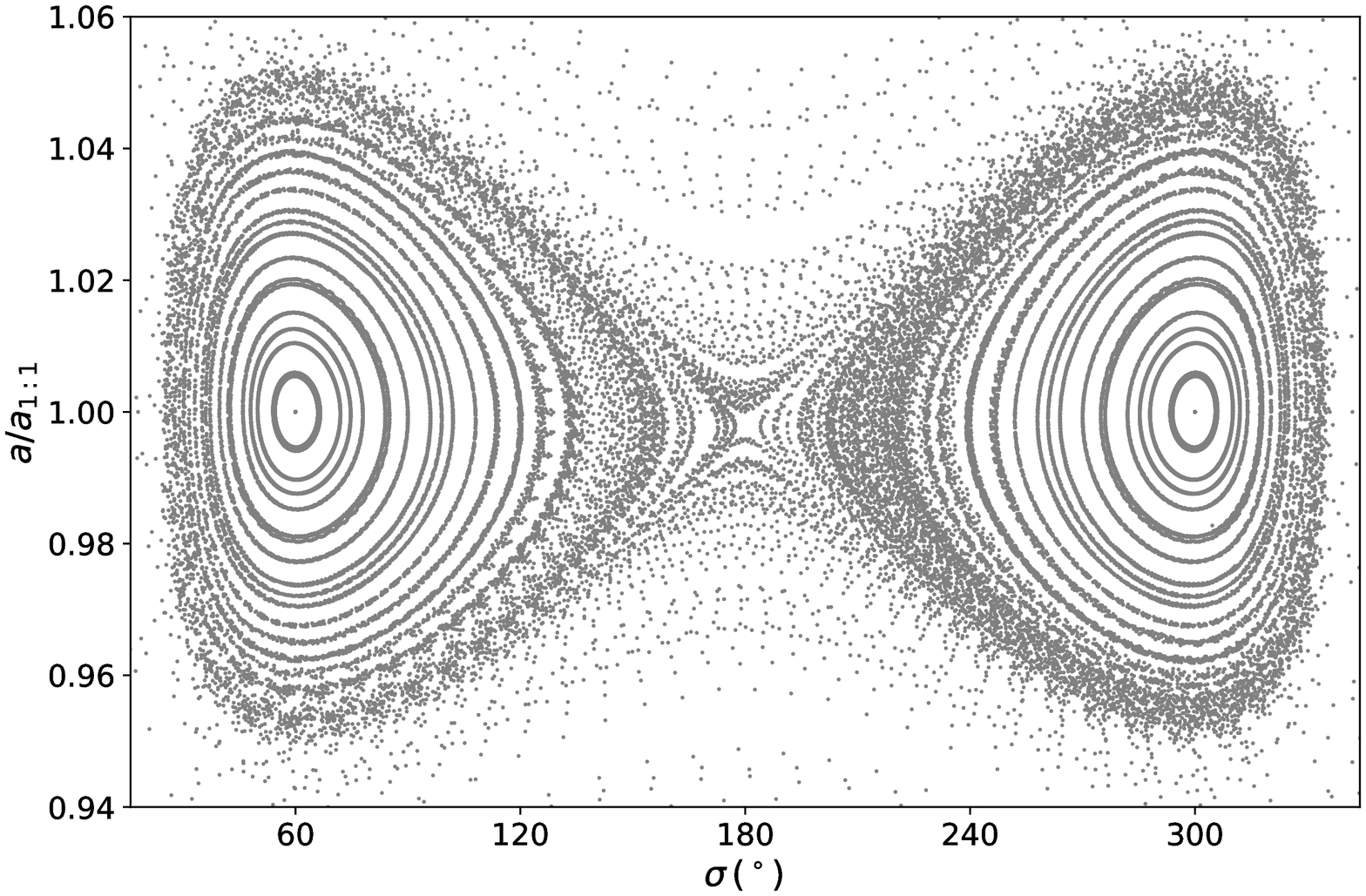}}
\resizebox{1.0\hsize}{!}{\includegraphics[width=\linewidth]{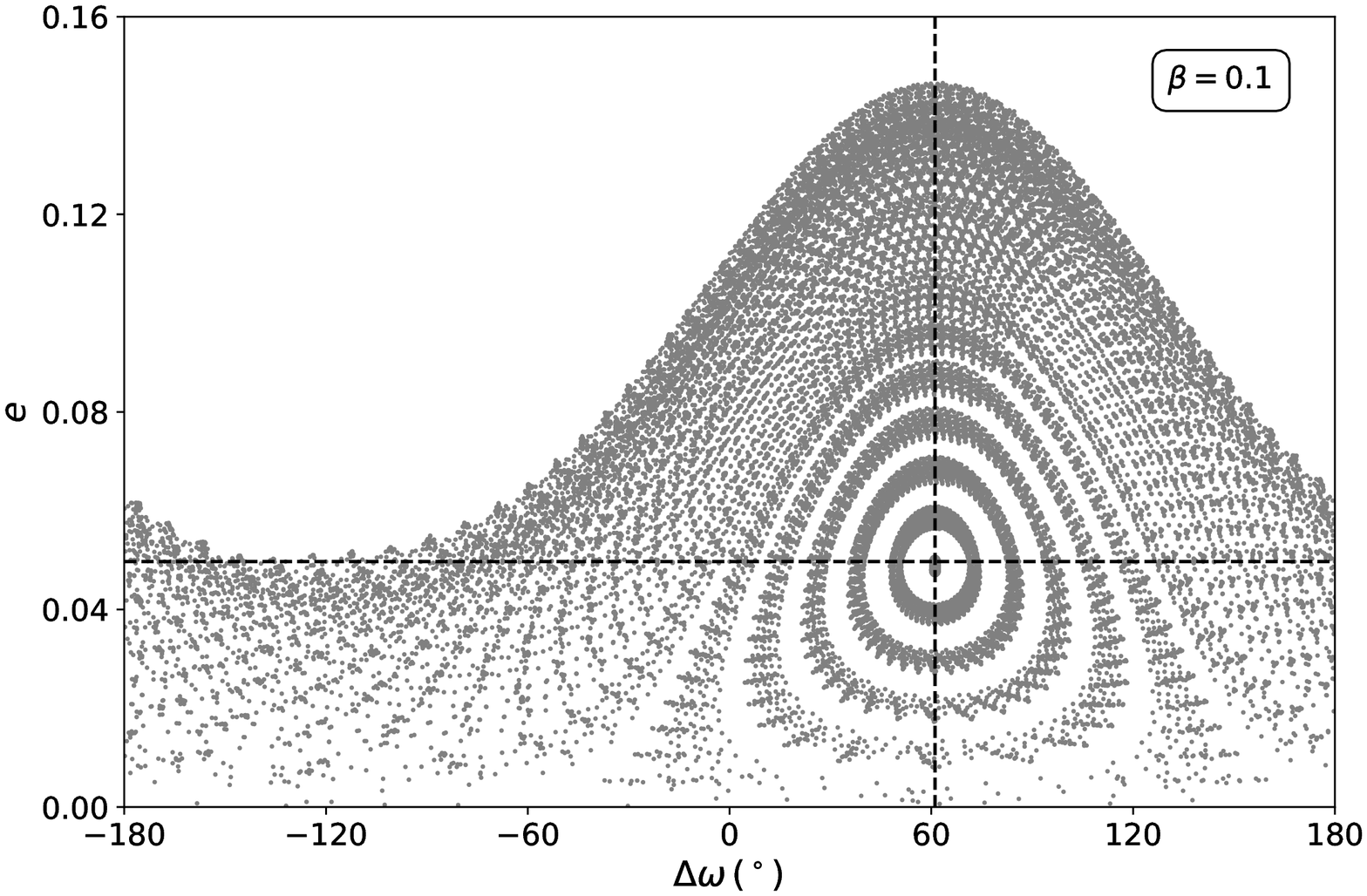}}
	\caption{Top: phase portrait (section $\sigma$ vs.
	$a/a_{1:1}$) for value of parameter $\beta=0$, $\gamma=0$. Bottom:
	projection ($\Delta\omega=\omega-\omega_J$ vs. $e$) for $\beta=0.1$
	and uncharged dust particles in the vicinity of $L_5$.}
\label{f:a-sigb0b}
\end{figure}

We start with a short phenomenological description of the phase space in
dependency of the system parameters. First, we integrate \equ{e:eom} and
choosing 60 initial conditions within the orbital plane of Jupiter with
$\sigma(0)$ (where $\sigma=\lambda-\lambda_J$ and mean orbital longitudes
$\lambda$ and $\lambda_J$) ranging from $0$ to $360^\circ$ and setting $\omega(0)$,
$\Omega(0)$ equal $\omega_J\pm60^\circ$, $\Omega_J$ as well as $a(0)$, $e(0)$
equal $a_J$, $e_J$. We stop the integration of the individual orbits at time
$t=5000$ yr or at close encounter with planet Jupiter.
 The simulation time has been fixed to ensure a complete
covering of the phase space. We notice, that erosion timescales of dust in space
may be much shorter, and strongly depend on the space environment and the chemical 
composition of the dust grain \citep[see, e.g.][]{2019SSRv..215...11S}.
The projection of the
orbits to the plane $(\sigma,a)$ for $\beta=0$ and $\gamma=0$ is shown in
Fig.~\ref{f:a-sigb0b}. In the classical problem we clearly see the location of
the tadpole and horseshoe type regimes of motion. The centers of librational
kinds of motions are located at $\sigma=\sigma_4=60^\circ$ and
$\sigma=\sigma_5=300^\circ$ -  that define the positions of the Lagrange points
$L_4$ and $L_5$, respectively. The saddle, denoted by $L_3$ is situated at
$\sigma=\sigma_3=180^\circ$ at the crossing of the separatrix that divides
librational and rotational motion close to resonance. We notice that for the
center orbits $e(t)\simeq e(0)$, $\omega(t)\simeq\omega(0)$, $\forall t$, and
that the uncharged dust grains stay within the orbital plane of Jupiter during
the whole integration time (not shown here). For the same set of parameters and 
initial conditions as before we integrate \equ{e:eom}, but with $\sigma(0)=\pm60^\circ$ 
and taking the values for $\Delta\omega(0)\in(-180^\circ,180^\circ)$, with 
$\Delta\omega=\omega-\omega_J$. We do not show the results for the classical
problem ($\beta=0$) here, but report that oscillatory behaviour takes
place close to $\Delta\omega=\pm60^\circ$, $e=e_J$, with increasing libration 
amplitudes for larger values of $e(0)$ as expected and already shown by
previous studies.\\ 

\begin{figure} \centering
\resizebox{1.0\hsize}{!}{\includegraphics[width=\linewidth]{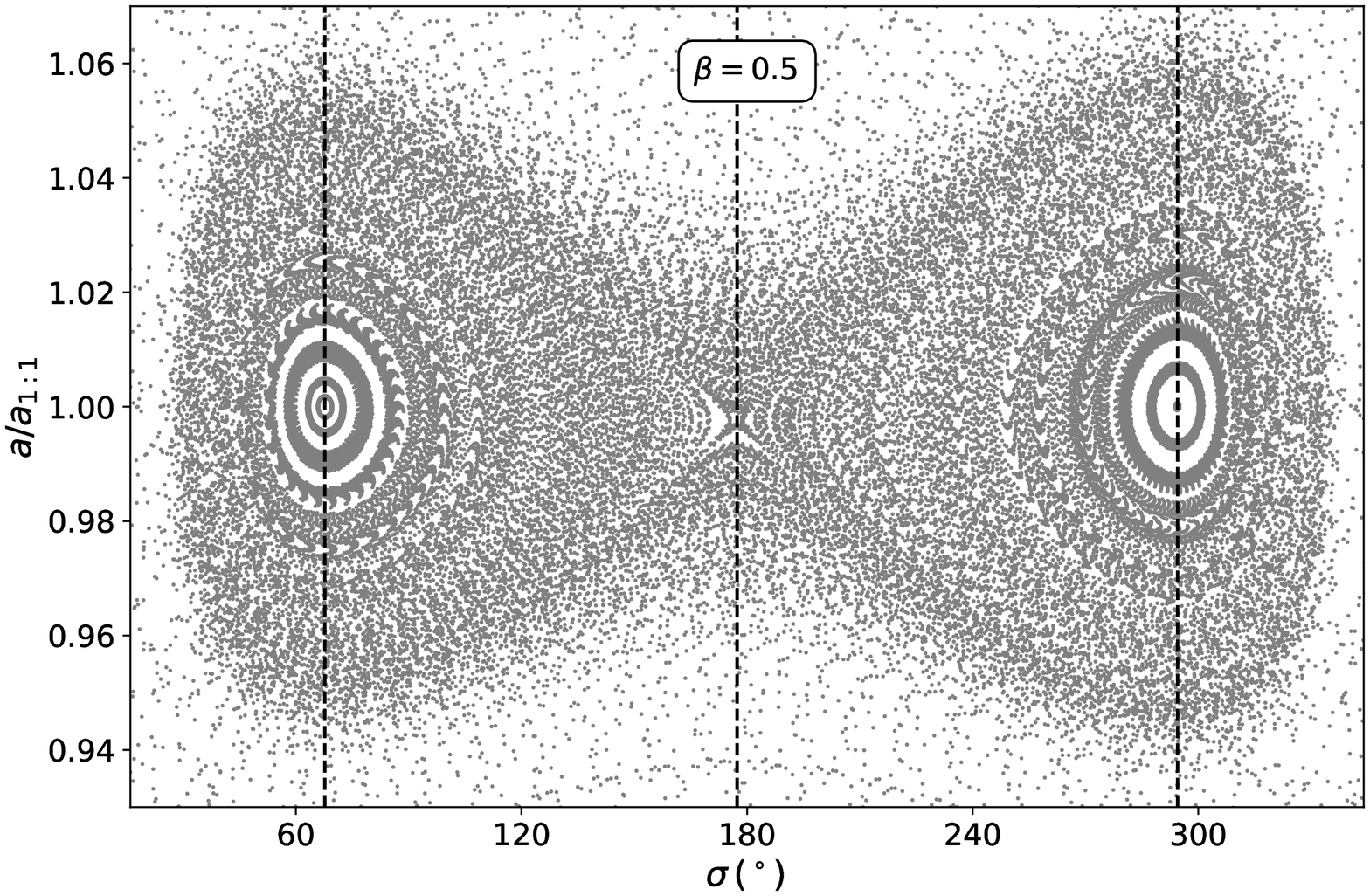}}
\resizebox{1.0\hsize}{!}{\includegraphics[width=0.5\linewidth]{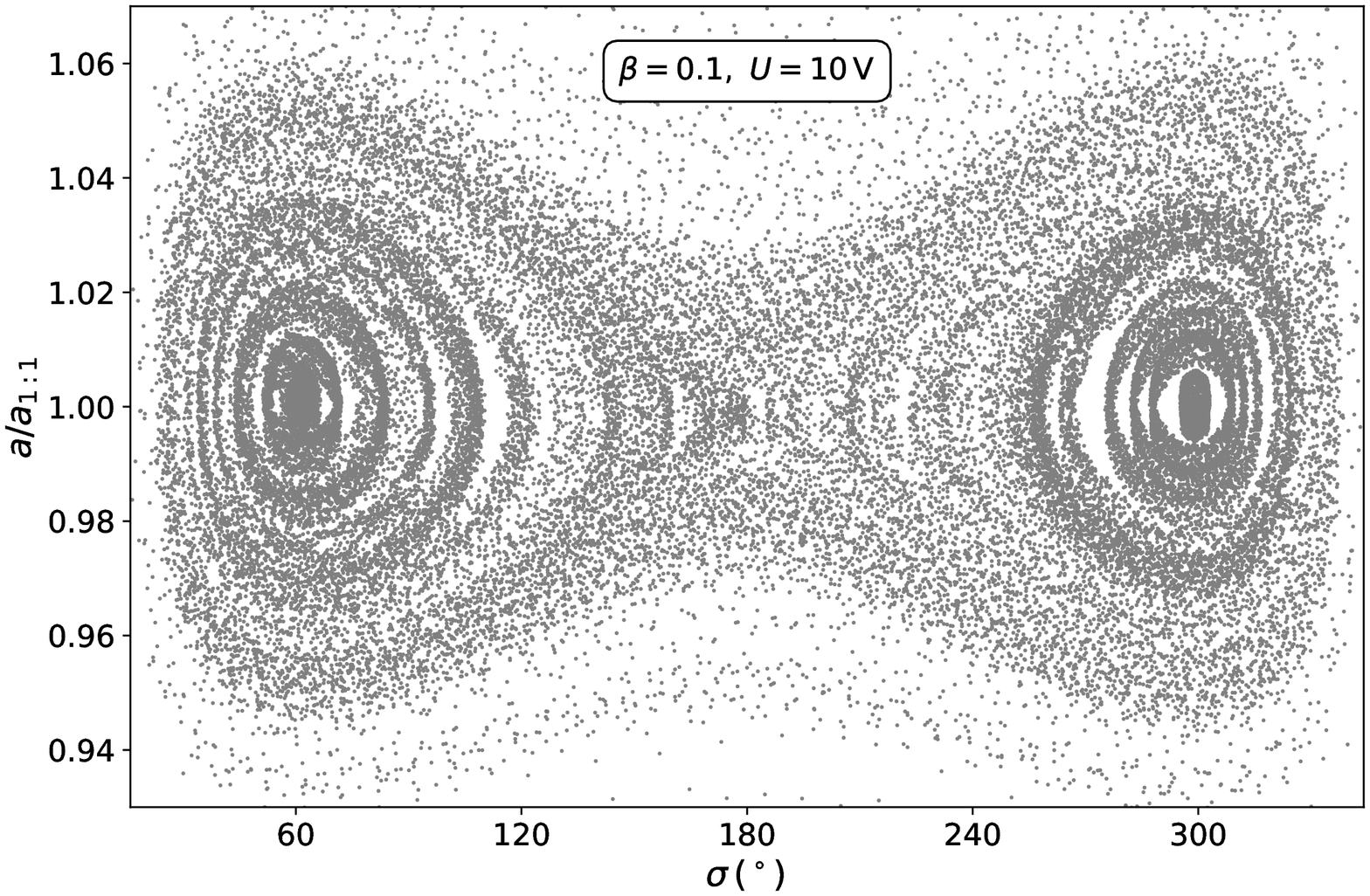}}
	\caption{Top: phase portrait (section $\sigma$ vs.
	$a/a_{1:1}$) for value of parameter $\beta=0.5$, $\gamma=0$. The resonant angles of
	the Lagrangian points $L_4$, $L_3$ and $L_5$, shown by vertical lines, are
	$67.71^\circ$, $177.34^\circ$ and $294.48^\circ$, respectively. Bottom:
	same initial conditions, but with $\beta=0.1$ and $10V$ surface potential.}
\label{f:a-sigb05a}
\end{figure}

For $\beta\neq0$ the perturbations due to solar radiation pressure and the
combined Poynting-Robertson effect and solar wind drag lead to additional
distortions of the orbits. We provide the results in the
$(\Delta\omega,e)$-plane at the bottom of Fig.~\ref{f:a-sigb0b}. We notice that
the orbits are actually not following invariant curves, but rather resemble the
projection of a $6D$ phase space spanned by the Kepler elements
$(a,e,i,\omega,\Omega,M)$ to a suitable choice of variables in $2D$.  We also
integrate \equ{e:eom} for 60 initial conditions up to integration time 5000 yr
and project the orbits to the plane $(\sigma,a)$.  The results are shown at the
top of Fig.~\ref{f:a-sigb05a} and should be compared to Fig.~\ref{f:a-sigb0b}.
While the orbits for the case $\beta=0$ follow the geometry of the classical
CRTBP  (Circular Restricted Three-Body Problem) with narrow widths close
to the libration centers, the majority of the orbits at the top in
Fig.~\ref{f:a-sigb05a}  fill up the full tadpole regime of motions.  The
asymmetry in the shifts from $60^o$, $-60^o$ between $L_4$, $L_5$ is clearly
present in the plane $(\sigma,a)$. While the offset from the location of $L_4$
from $60^o$ is about $7.71^o$ for $\beta=0.5$ the shift is only about $5.52^o$
from the location of $L_5$ at $300^o$ for $\beta=0$. As it can be seen by the
vertical dashed lines in Fig.~\ref{f:a-sigb05a} the shift from $180^o$ (the
location of $L_3$ for $\beta=0$) is about $2.66^o$ at $\beta=0.5$. This
asymmetry in phase space due to the different response of micron-sized
particles between $L_4$ and $L_5$ will be addressed in full detail later in
this section.  \\

\begin{figure} \centering
\resizebox{1.1\hsize}{!}{\includegraphics[width=\linewidth]{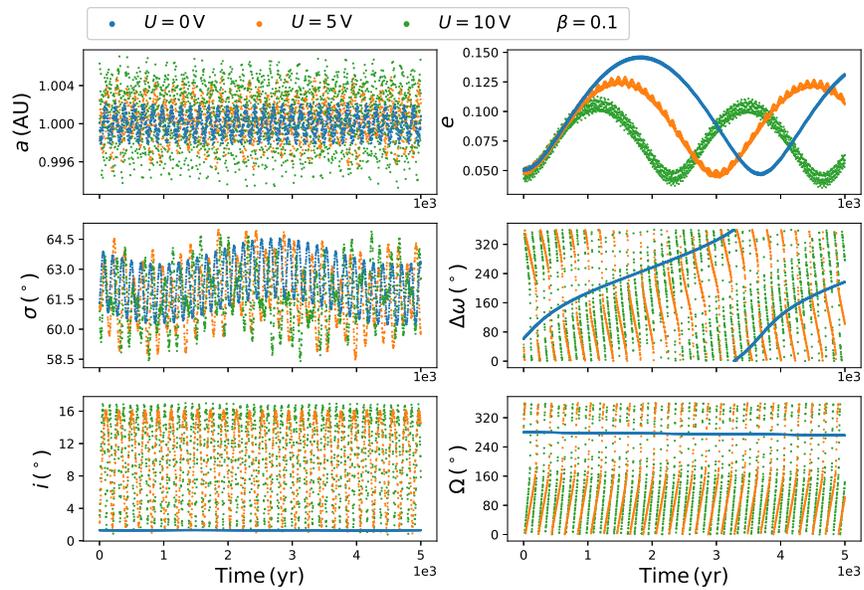}}
	\caption{Orbital evolution of particles starting from the location of
	$L_4$ (but $\Omega(0)=\Omega_J+180^o$) in the uncharged problem with
	$\beta=0.1$. The  blue curve indicates the uncharged model while the
	orange and green curves indicate the orbital evolution of charged
	particles.}
\label{f:XH}
\end{figure}

The role of charge on the dynamics is three-fold: i) first, the additional
perturbations lead to even stronger distortions when projected to the plane
$(\sigma,a)$ - see bottom of Fig.~\ref{f:a-sigb05a}, where 30 orbits with
initial $\sigma(0)$ from 0 to $360^\circ$ are integrated for 5000 yr; ii) the
interaction with the interplanetary magnetic field prevents the charged dust
particles to perform regular, oscillatory type of motions in the
$(\Delta\omega,e)$-plane (but still close to the centers, not shown here); iii)
finally, Lorentz force acts transversal to the orbital plane of Jupiter and
leads to excursions of the dust grains to high ecliptic latitudes that cannot be
seen in the uncharged problem, see lower left of Fig.~\ref{f:XH}, where we show
three individual orbits starting with the same initial conditions, but
different charge-to-mass ratios corresponding to $0V$, $5V$, and $10V$ surface
potential, respectively.\\

The comparison of the pure gravitational problem together with the uncharged one
including solar radiation pressure and Poynting-Robertson effect with the charged
problem, already demonstrates the role of $\beta$ and $\gamma$ on the
topology of the phase space. Next, we aim to quantify the dependency on these
system parameters of 1) the location of the orbit with minimum libration
amplitudes (that corresponds to an equilibrium in the classical problem), 2)
the effect of $\beta$ and $\gamma$ on the libration width of the 1:1MMR, and 3)
their role on the time of temporary capture close to resonance. To start with
1) we require a robust and suitable numerical tool to be defined in the next section.

\subsection{Minimum libration amplitude solutions}

Let $w_T=w_T(X_j,\vec X_0;\beta;\gamma)$ be the maximum libration amplitude of the
$j$-th component $X_j=X_j(t)$ of vector $\vec X=\vec X(t)$ during the
integration time $0\leq t\leq T$, starting with the initial condition $\vec
X_0=(a_0,e_0,i_0,\Delta\omega_0,\Omega_0,\sigma_0)$, and for fixed values of
the parameters $\beta$, $\gamma$. Let $\vec
X^*=(a^*,e^*,i^*,\Delta\omega^*,\Omega^*,\sigma^*)$ be the initial
condition that defines the location of $L_4$ with $a^*=a_J$, $e^*=e_J$,
$i^*=i_J$, $\Delta\omega^*=60^\circ$, $\Omega^*=\Omega_J$, and $\sigma^*=60^\circ$ (or
$\Delta\omega^*=300^*$, $\sigma^*=300^\circ$ in case of $L_5$). For vanishing
values of the parameters $\beta$, $\gamma$ and in exact steady-state
configuration we have $w_T(X_j,\vec X^*,0,0)\simeq0$ for all $j$, while for
non-zero $\beta$, $\gamma$ we generally find $w_T(X_j,\vec
X^*,\beta,\gamma)=\delta_j$ with $\delta_j>0$ and $j=1,\dots,6$. To obtain a
minimum libration amplitude solution we minimize $w_T$ with respect to $X_j$
starting in the vicinity of $\vec X^*$, say $\vec X_{\varepsilon_j}^*$ 
with $X_j(0)=X_j^*\pm\varepsilon_j$ and positive $\varepsilon_j$:

\beq{e:wT}
\delta_j=\min_{X_j^*-\varepsilon_j\leq X_j\leq X_j^*+\varepsilon_j}
w_T(X_j,\vec X_{\varepsilon_j}^*,\beta,\gamma), \quad j=1,\dots,6.
\eeq

\noindent Our aim is to minimize all $\delta_j$ with $j=1\dots6$ for fixed pair
$(\beta,\gamma)$.  We start with the case $\beta>0$ and $\gamma=0$. Since solar
wind drag and Poynting-Robertson effect does not influence ascending node
longitude and inclination of the orbital plane we keep $X_5=\Omega^*$, $X_3=i^*$ fixed.
Taking into account that the libration amplitudes are strongly coupled in pairs
$(a,\sigma)$ and $(e,\Delta\omega)$ - see phase portraits in the previous
section - we minimize \equ{e:wT} in an iterative way as follows. \\

Let $\vec X_{\varepsilon_j,r}^*$ be the initial condition during the $r$-th
iteration process with $\vec X_{\varepsilon_j,0}^*=\vec X_{\varepsilon_j}^*$.
We start with a set of initial conditions on a grid in $(a,\sigma)$ using
$a(0)=a_J(1-\beta)^{1/3}\pm\varepsilon_1$ and
$\sigma(0)=\sigma^*+\varepsilon_6$ with $\varepsilon_1>0$ and
$0\leq\varepsilon_6\leq 120^\circ$ and integrate \equ{e:eom} up to time
$T$.
 Here, the choice for $a(0)$ is motivated by the estimate that follows. Kepler's
3rd law for planet Jupiter takes the form:
\beqno
n_J^2a_J^3=\mu ,
\eeqno
while the law for the dust particle, including radiation pressure, becomes:
\beqno
n^2a^3=\mu(1-\beta) .
\eeqno
Eliminating $\mu$ in above equations we find the relation
\beqno
n^2a^3=n_J^2a_J^3(1-\beta) ,
\eeqno
and taking into account $n_J=n$, in presence of a 1:1 MMR, we finally arrive at
\beqno
a=a_J(1-\beta)^{1/3} .
\eeqno

Next, we choose, out of the set of orbits, the initial condition with the minimum
libration amplitude solution and identify it with $\vec
X_{\varepsilon_{1,6},r=1}^*$ and repeat the above iteration step with smaller
$\varepsilon_1$, $\varepsilon_6$ to obtain $\vec X_{\varepsilon_{1,6},r+1}^*$.
The iteration stops if no significant decrease in libration amplitude can be
found anymore when decreasing $\varepsilon_1$, $\varepsilon_6$.  Let $a^{**}$,
$\sigma^{**}$ be the initial condition of the minimum libration amplitude
solution at the final iteration step $R$. As it turns out the choice
$a(0)=a^{**}=a_J(1-\beta)^{1/3}$ is the correct choice for the minimum
libration amplitude solution for $\gamma=0$ while $\sigma(0)=\sigma^{**}$
is shifted from the equilibrium of the classical problem with increasing 
value of $\beta$. Next, we repeat above procedure by
fixing $a(0)=a^{**}$, $\sigma(0)=\sigma^{**}$, and minimizing $w_T$ with 
respect to $X_2(0)=e(0)$, $X_4(0)=\Delta\omega(0)$ on a grid $(e,\Delta\omega)$
and using $\varepsilon_2$, $\varepsilon_4$.  We start by
integrating \equ{e:eom} with initial conditions $e(0)=e_J\pm\varepsilon_2$ and
$\Delta\omega(0)=\pm60\pm\varepsilon_4$ with $\varepsilon_2, \varepsilon_4>0$,
and again identify the initial condition 
that results in the minimum libration amplitude solution with 
$\vec X_{\varepsilon_{2,4}=0,r=1}$. We repeat the iterative procedure
to obtain $e^{**}$ and $\Delta\omega^{**}$ at iteration step $r=R$.
As it turns out, for the case $\gamma=0$, $w_T$ is minimal for the
choice $e(0)=e^{**}=e_J$, while
$\Delta\omega^{**}$ gets shifted from the equilibria
defined for $\beta=0$ with increasing values of $\beta$ like in
the case for $\sigma$.

The whole process is done for different values of
$\beta$ and for starting values close to $L_4$ and also $L_5$.  Let us denote
by $a_k$, $e_k$, $i_k$, $\Delta\omega_k$, $\Omega_k$, and $\sigma_k$ the final
optimal values that minimize $w_T$, with $k=4$ in the vicinity of $L_4$ and
$k=5$ close to $L_5$ for fixed value of $\beta$. \\

\begin{figure} \centering
\resizebox{1.0\hsize}{!}{\includegraphics[width=\linewidth]{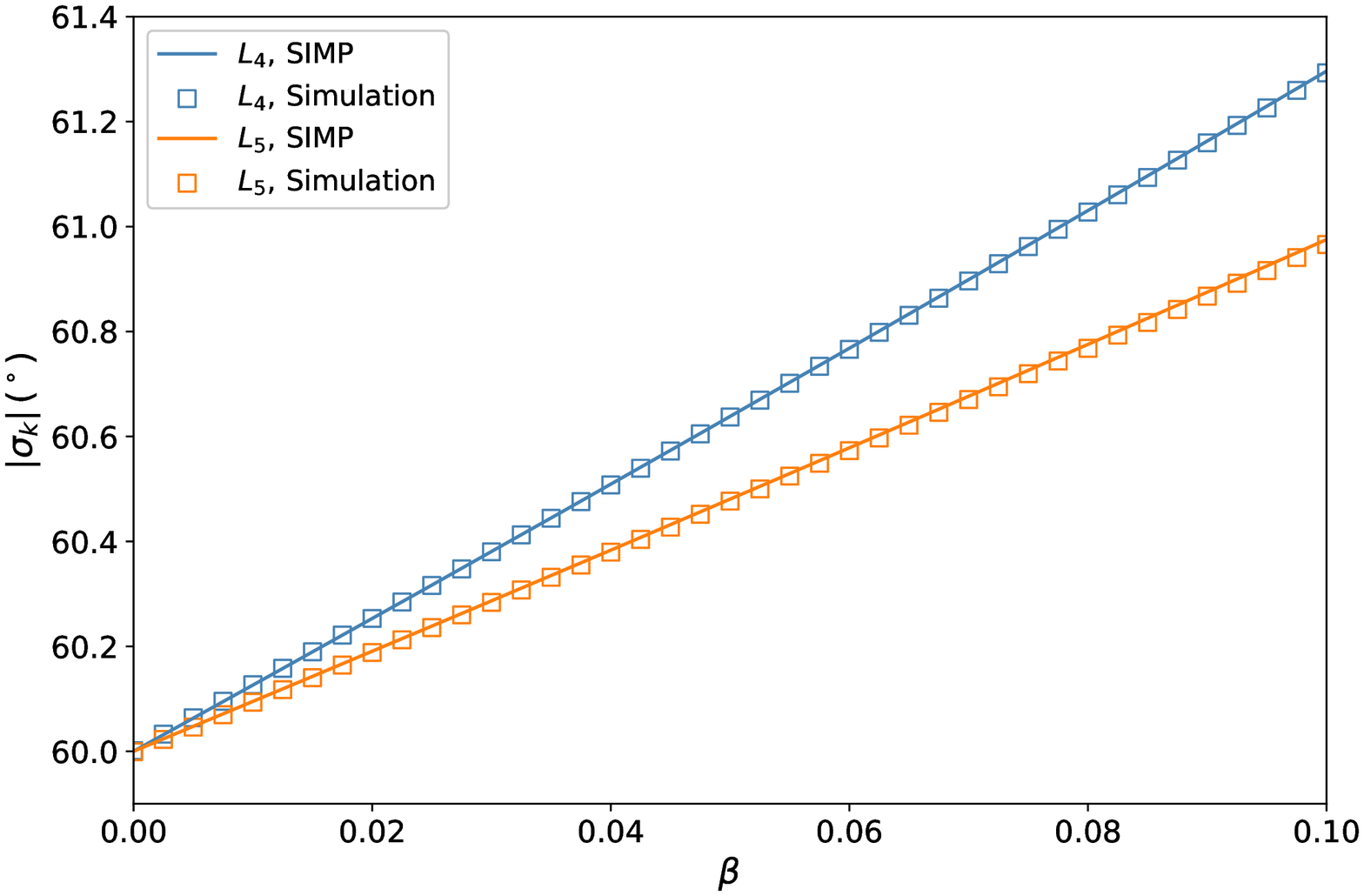}}
\resizebox{1.0\hsize}{!}{\includegraphics[width=\linewidth]{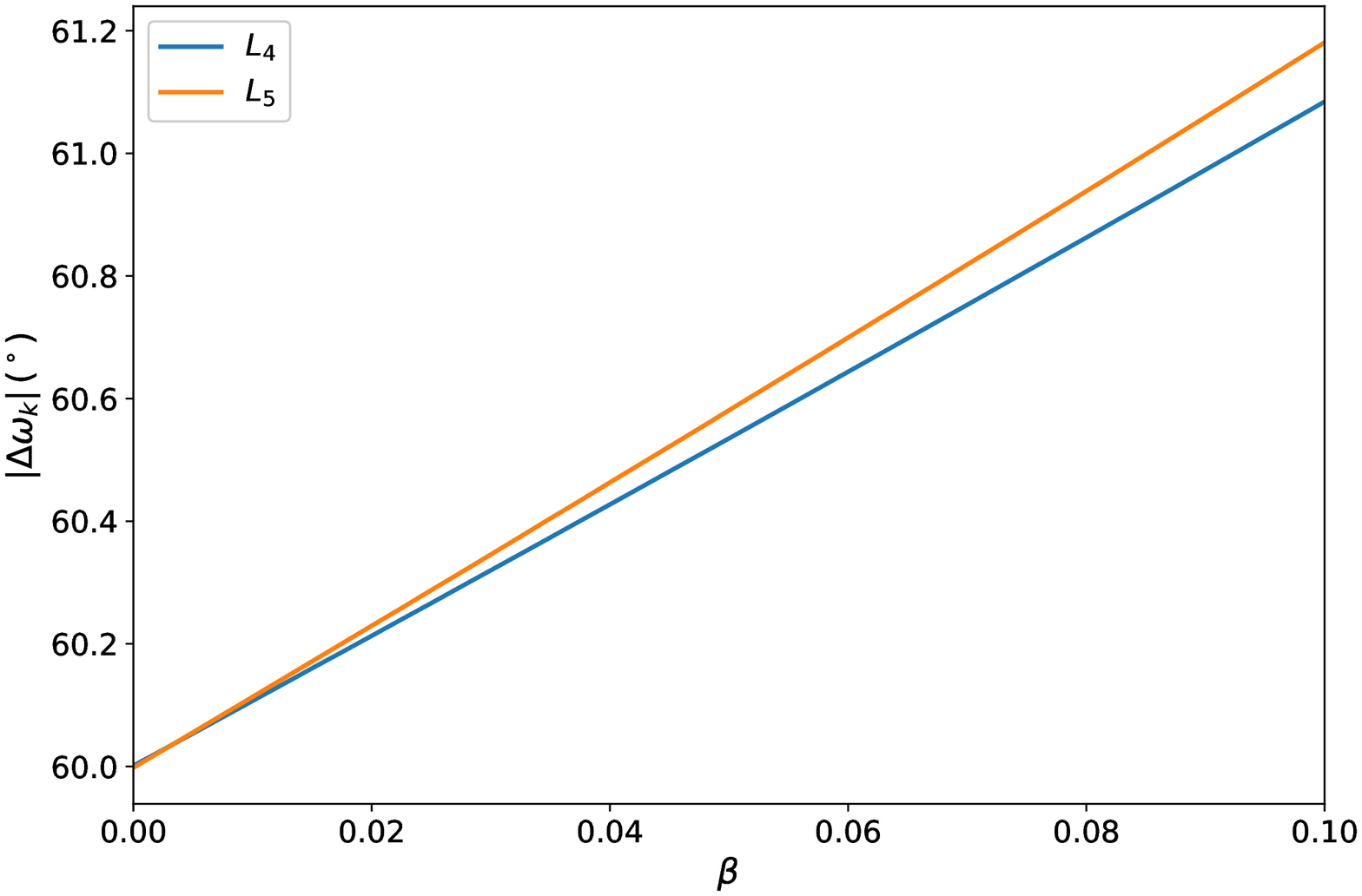}}
	\caption{Dependency of minimum libration amplitude in 
	$\sigma_k$ (top) and $\Delta\omega_k$ (bottom) with $k=4,5$ corresponding to $L_k$ 
	on parameter $\beta$ (and $\beta=0$ indicated by black-dashed line). We note that the $y-$axis is the absolute value of $\sigma_k$ and $\Delta\omega_k$ in the range of $(-180^\circ,180^\circ)$.}
\label{f:sigdw-45}
\end{figure}

The results for variables $\sigma_k$ and $\Delta\omega_k$ are shown in 
Fig.~\ref{f:sigdw-45}. On the top we report the value of $\sigma$ 
(by squares) with the minimum of the maximum libration amplitude during the
integration time for given values of the parameter $\beta$ in the 
vicinity of $L_4$ (blue) and close to $L_5$ (orange). The lines are obtained
from a theory based on the circular restricted three-body problem including 
solar radiation pressure and the Poynting-Robertson effect 
\citep[see, e.g.][]{1980ApJ...238..337S, 2021A&A...645A..63Z} and confirm the values
obtained by minimizing \equ{e:wT}. At the bottom of Fig.~\ref{f:sigdw-45}
we report the results for $\Delta\omega_k$ close to $L_4$ (blue) and
$L_5$ (orange), both obtained by minimizing \equ{e:wT}. In both figures 
we clearly see the shift of the minimum libration amplitude solutions 
in dependency of parameter $\beta$ up to $1.5$ degrees at $\beta=0.1$.
The two panels in Fig.~\ref{f:sigdw-45} also reveal an asymmetry 
between the Lagrange points $L_4$ and $L_5$. While the shift in $\sigma$ from
$60^\circ$ is more prominent for solutions close to $L_4$, in comparison to
the shift from $-60^\circ$ (close to $L_5$), the situation is reversed in the 
shift of $\Delta\omega$. The findings confirm the results, already  obtained in
\cite{2015Icar..250..249L}, where the authors use a semi-analytical approach
in the framework of the restricted three-body problem (circular, elliptic, and spatial)
that is based on averaged equations of motions. \\

\begin{figure}
	\centering
    \resizebox{1.0\hsize}{!}{\includegraphics{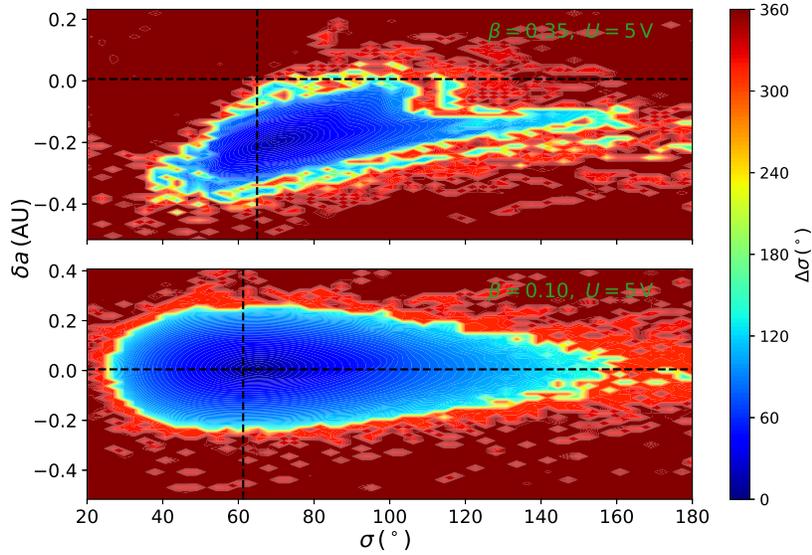}}
		\caption{Libration amplitudes
		($\Delta\sigma$) on the $(\sigma,a)$ plane in the charged
		problem ($U=5$ V) with $\beta=0.1$ (bottom) and $\beta=0.35$
		(top). The dashed lines indicate $(\sigma_k,a_k)$ in the
		uncharged problem for the corresponding value of $\beta$. We
		note the $y$-axis is the deviation from the nominal value $a_{1:1}$.}
    		\label{f:dsigV}
	\end{figure}

Next, we repeat by minimizing \equ{e:wT} wrt. $a(0)$, $e(0)$,
$\Delta\omega(0)$, $\sigma(0)$ for $\gamma>0$, i.e. for dust grain surface
potentials equal $5$ and $10$ Volts, in the same way as for the case
$\gamma=0$. We find a strong influence of charge on the dynamics as
demonstrated by the example shown in Fig.~\ref{f:dsigV}. The figure shows the
libration amplitudes $\Delta\sigma$ on a grid of initial conditions
$\sigma(0)\times \delta a(0)$, with $\delta a = a-a_{1:1}$, for two different
values of $\beta$ and $5$ Volts surface charge in the vicinity of $L_4$. The
location of the minimum libration amplitude solutions in the uncharged case are
given by black-dashed lines and we clearly see that in the presence of charge
this location (dark-blue) gets shifted towards smaller values in semi-major
axis $a$ and larger values in $\sigma$.  The effect becomes stronger for larger
values of $\beta$ (allowing larger values in $\gamma$ for same surface
potential charge): while the deviation in $\delta a$ is smaller and symmetric
with respect to the symmetry line $\delta a=0$ it becomes larger and asymmetric
in the case $\beta=0.35$. We also notice that with increasing $\beta$ the
region of librational motion shrinks. Solar wind, Poynting-Robertson effect as
well as the interplanetary magnetic field strongly affect the dynamics.
In the following we report the location and parameters for the minimized
solution with respect to librational amplitudes, i.e. the location of the 
darkest blue region in simulations of the type as shown in Fig.~\ref{f:dsigV},
in dependency of the system parameters.

\begin{figure} \centering
\resizebox{1.0\hsize}{!}{\includegraphics[width=\linewidth]{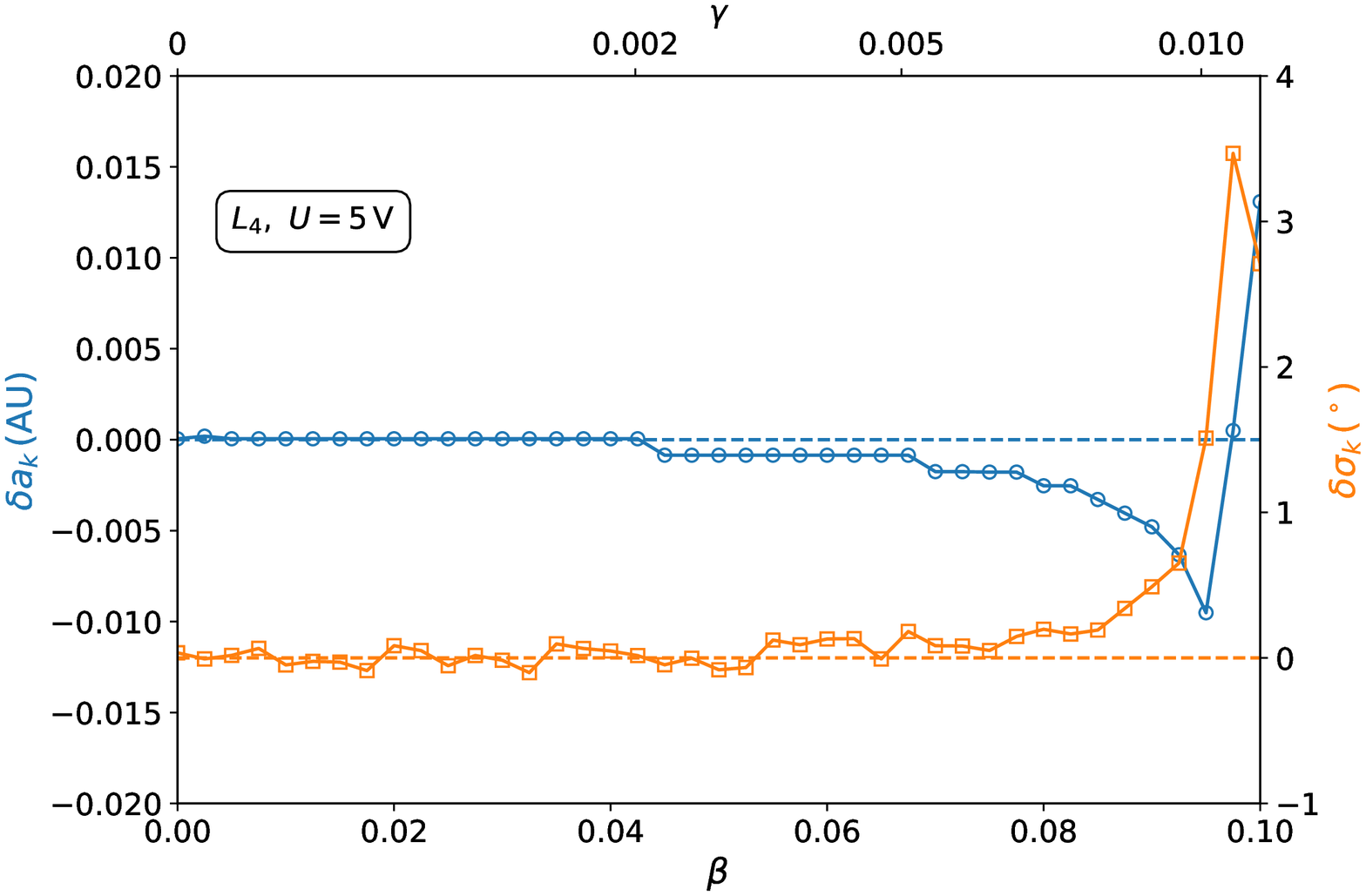}}
\resizebox{1.0\hsize}{!}{\includegraphics[width=\linewidth]{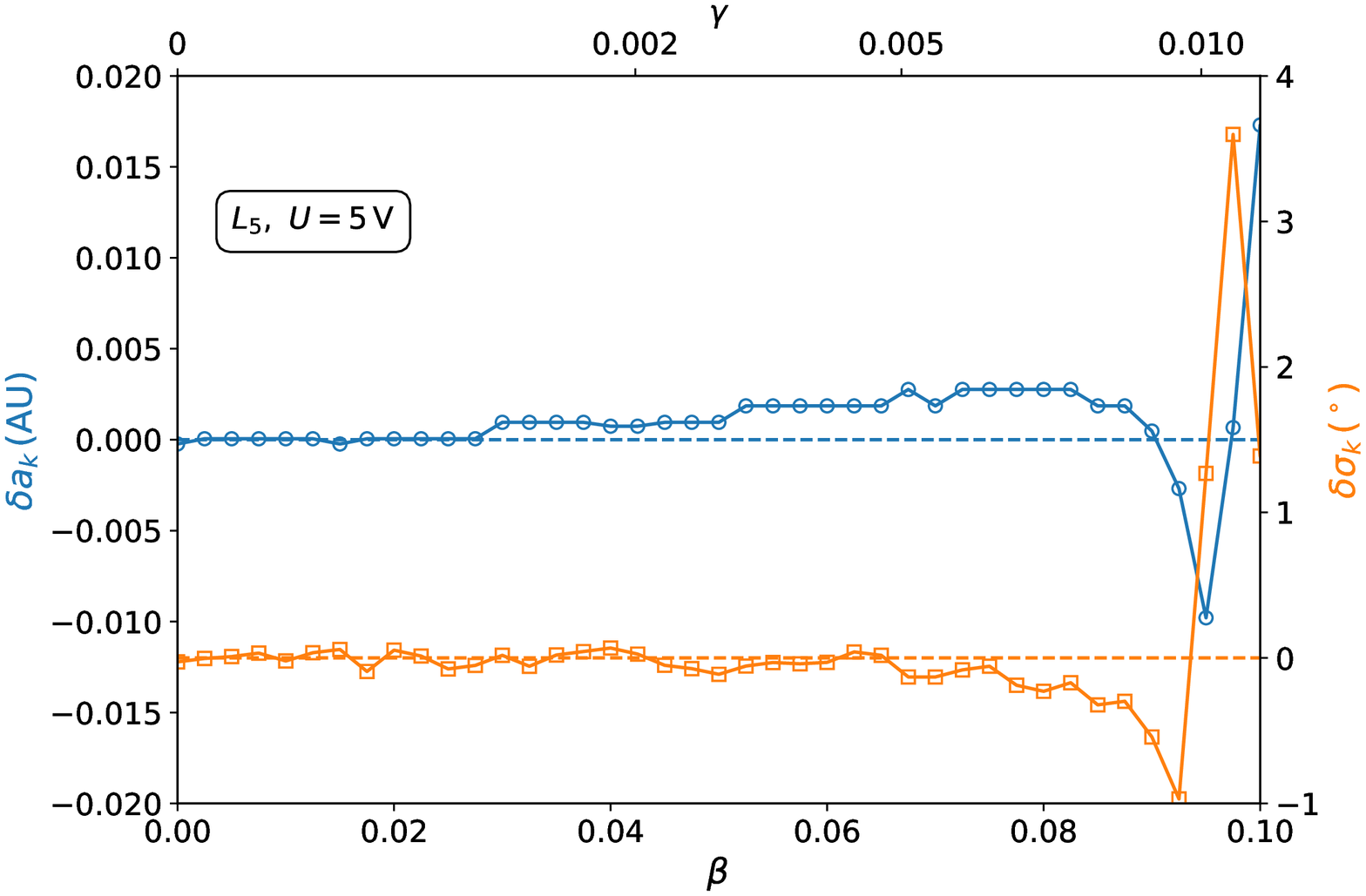}}
	\caption{Deviation from minimum libration amplitudes of
	uncharged problem for dust grain potential surface charge of $5$ Volts in
	semi-major axes $\delta a_k$ and resonant angles $\delta\sigma_k$ (with $k=4$
	in the vicinity of $L_4$ - top and $k=5$ close to $L_5$ - bottom). The cases with 
	$V=0$ Volts are shown by dashed lines.}
\label{f:Delta-asig-kV}
\end{figure}

\begin{figure} \centering
\resizebox{1.0\hsize}{!}{\includegraphics[width=\linewidth]{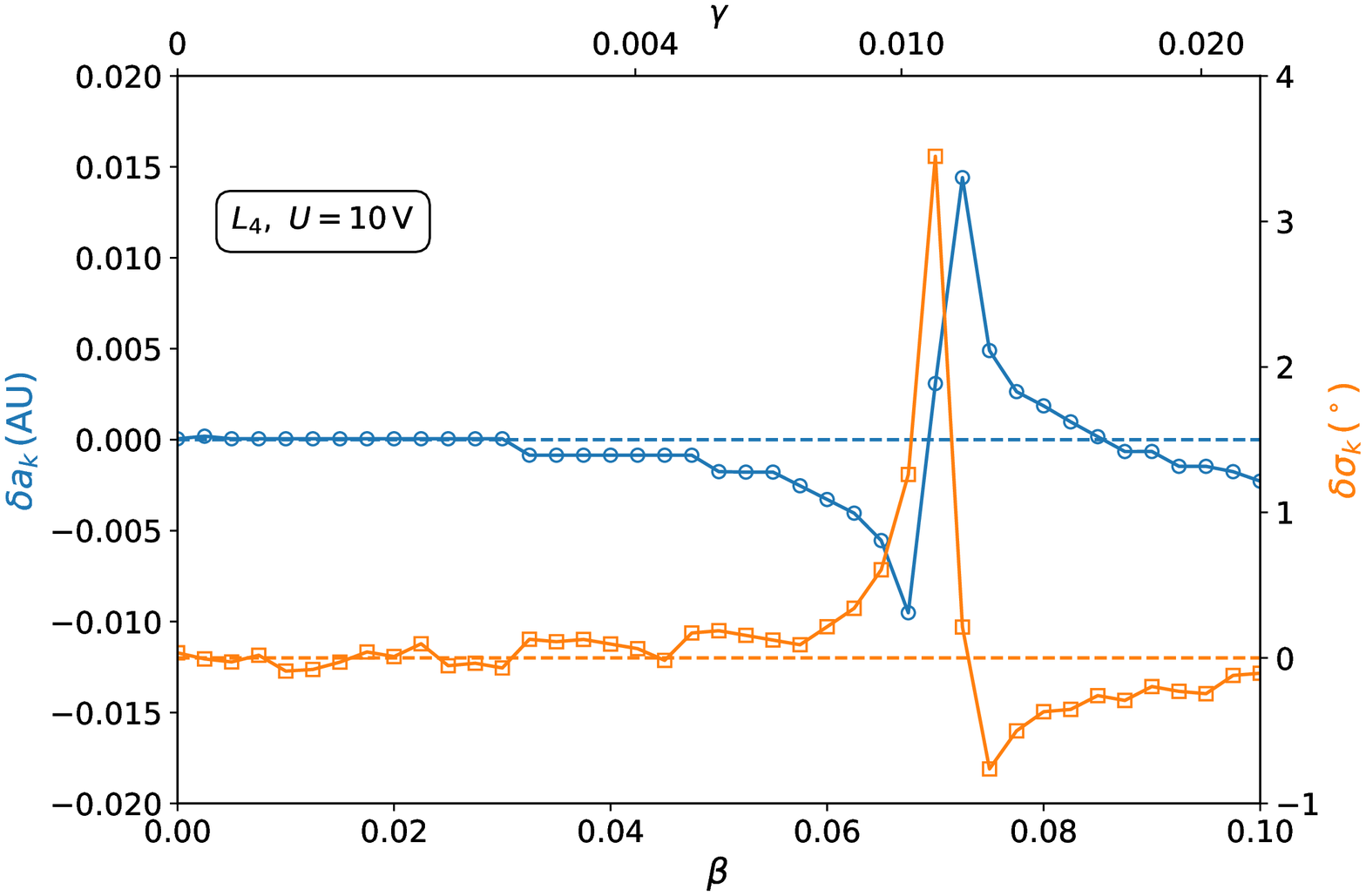}}
\resizebox{1.0\hsize}{!}{\includegraphics[width=\linewidth]{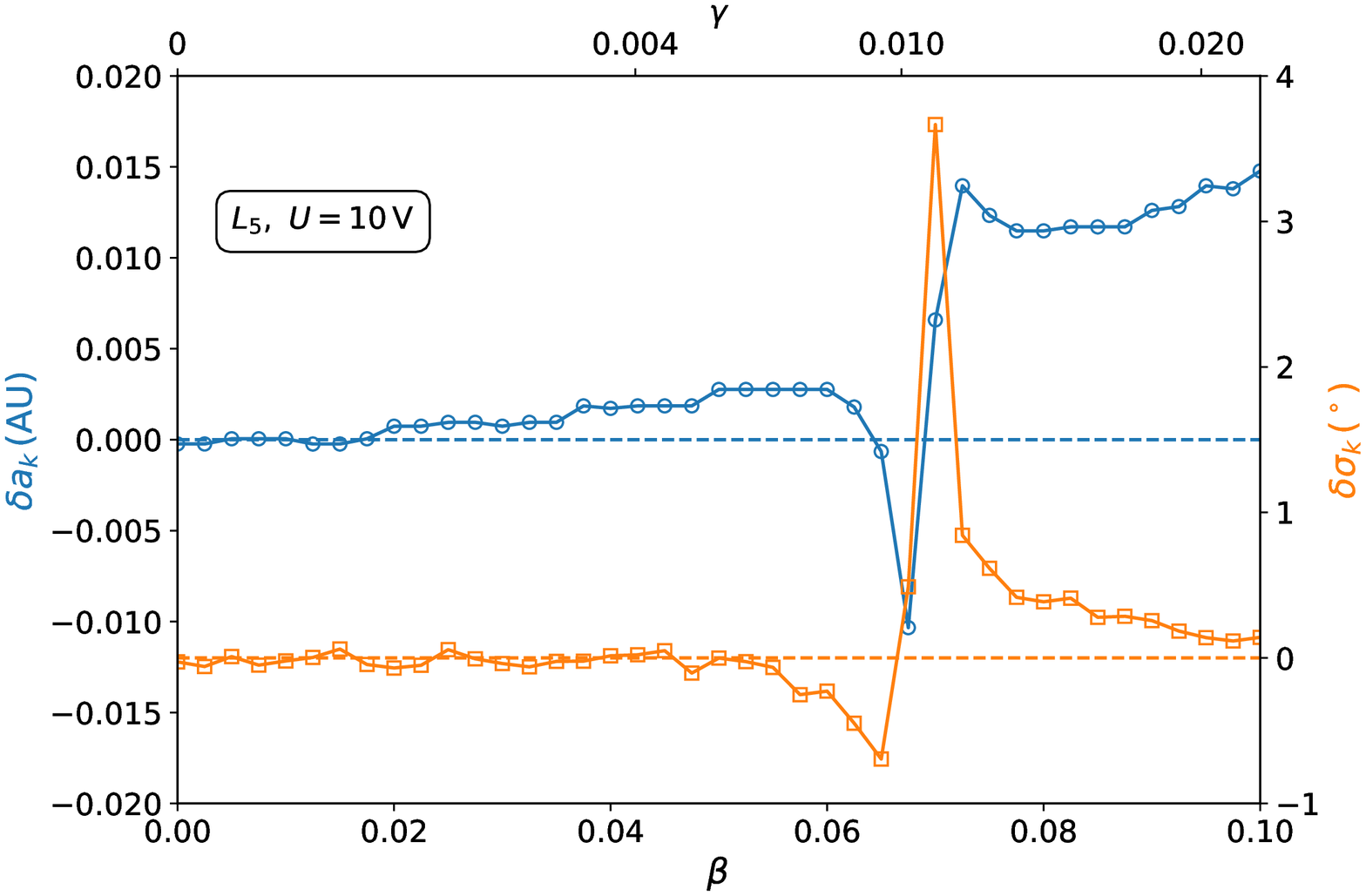}}
	\caption{Deviation from minimum libration amplitudes of
	uncharged problem for dust grain potential surface charge of $10$ Volts in
	semi-major axes $\delta a_k$ and resonant angles $\delta\sigma_k$ (with $k=4$
	in the vicinity of $L_4$ - top and $k=5$ close to $L_5$ - bottom). The cases with 
	$V=0$ Volts are shown by dashed lines. Compare with Fig.~\ref{f:Delta-asig-kV}.}
\label{f:Delta-asig-kX}
\end{figure}

We summarize our study for different values of $\beta$ in case of dust grain surface
potentials equal $5$ and $10$ Volts in
Fig.~\ref{f:Delta-asig-kV},~\ref{f:Delta-asig-kX}, where the results  are given
relative to the solution of minimum libration amplitude of the uncharged case.
At the top of Fig.~\ref{f:Delta-asig-kV} we report the shift $\delta a_4$ in
blue and relative to $a_4$ due to $\gamma>0$ for $5$ Volts. With increasing
value of $\beta$ we find a negative shift that becomes larger in magnitude and
reaching 0.01 AU for $\beta=0.1$.  In Fig.~\ref{f:Delta-asig-kV} we also report
the shift $\delta\sigma_4$ in orange and relative to $\sigma_4$, that again
becomes larger with increasing values of $\beta$ (and $\gamma$) reaching several
degrees of offset at $\beta=0.1$. The results in $\delta a_5$, $\delta \sigma_5$ are 
shown at the bottom of Fig.~\ref{f:Delta-asig-kV} with comparable values in 
$\delta a_5$ ($\delta \sigma_5$) in comparison with $\delta a_4$ ($\delta \sigma_4$). 
We notice the presence of spikes close to $\gamma\simeq0.01$ in both panels that 
are also present for different surface potential of the dust grains 
(see, Fig.~\ref{f:Delta-asig-kX} in case of $10$Volts). We notice that for larger
values of the surface potential these spike like structure is shifted towards 
smaller values in $\beta$ but still remains close to the value $\gamma\simeq0.01$.

\begin{figure}
	\centering
    \resizebox{1.1\hsize}{!}{\includegraphics{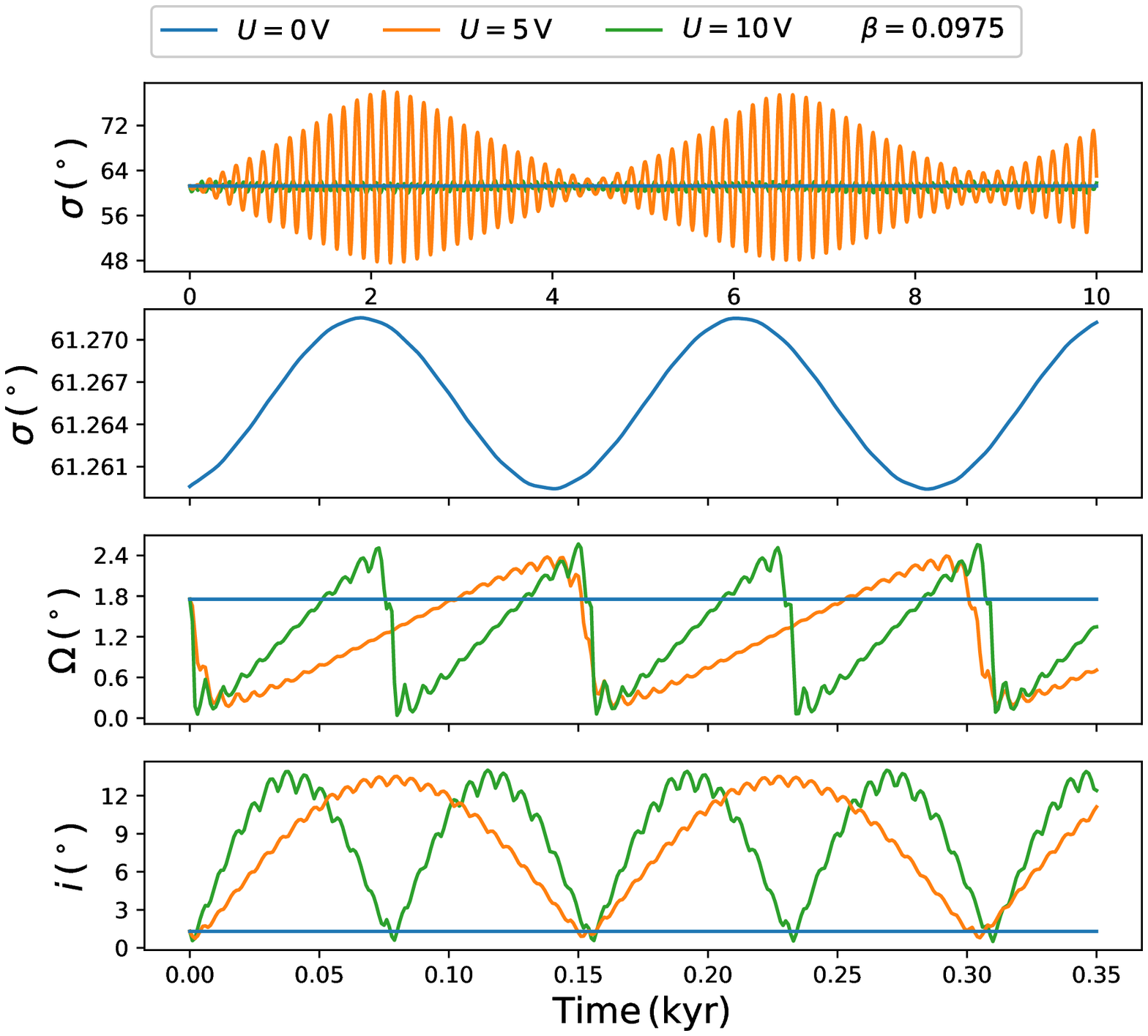}}
		\caption{Occurence of the spike in 
		Fig.~\ref{f:Delta-asig-kV} for $\beta\simeq0.1$, 
		$\gamma\simeq0.01$ (orange with $U=5V$). Enhancement
		of amplitude variations of resonant argument, resonant argument
		in uncharged case ( blue with $U=0V$), evolution of ascending node 
		longitude, and inclination (from top to bottom). For $U=10V$ (green)
		the spike does not show up at $\beta\simeq0.01$.}
    		\label{f:spikes}
	\end{figure}

Where does it come from? A series of simulations with various initial conditions
and parameters suggests that the spikes are the result of a commensurability between
the precession rate $\dot\Omega$ of the ascending node longitude and the period 
in libration of the resonant angle $\sigma$. As demonstrated in Fig.~\ref{f:spikes}
the variations in amplitude of the angle $\sigma$ are greatly enhanced for the case
$U=5V$ (in orange), where the fundamental period in $\Omega$ (and $i$) is about the 
same as for the resonant argument itself. On the contrary, the amplitude variations
stay small during the whole integration period in the uncharged case $U=0V$, and for 
the case $U=10V$. We remark that the period in $\Omega$ is inversely proportional to the 
parameter $\gamma$, or $\Gamma$ - as it has been shown in \citet{2019CeMDA.131...49L}.
As a result, for $U=10V$, the period of $\Omega$ and $i$ is only half of the value for
$U=5V$, as it is confirmed also by the bottom two panels of Fig.~\ref{f:spikes}.
Since the libration period of $\sigma$ stays the same for different
values of $\beta$, the values of the  parameter $\gamma$ which determine the period
of $\Omega$ could finally determine the locations of the spike at the same value of
$\gamma$, although the corresponding values of $\beta$ are different, since 
$\beta\propto 1/R$ while $\gamma\propto U/R^2$ - see also \equ{e:betgam}.

\begin{figure} \centering
\resizebox{1.0\hsize}{!}{\includegraphics[width=\linewidth]{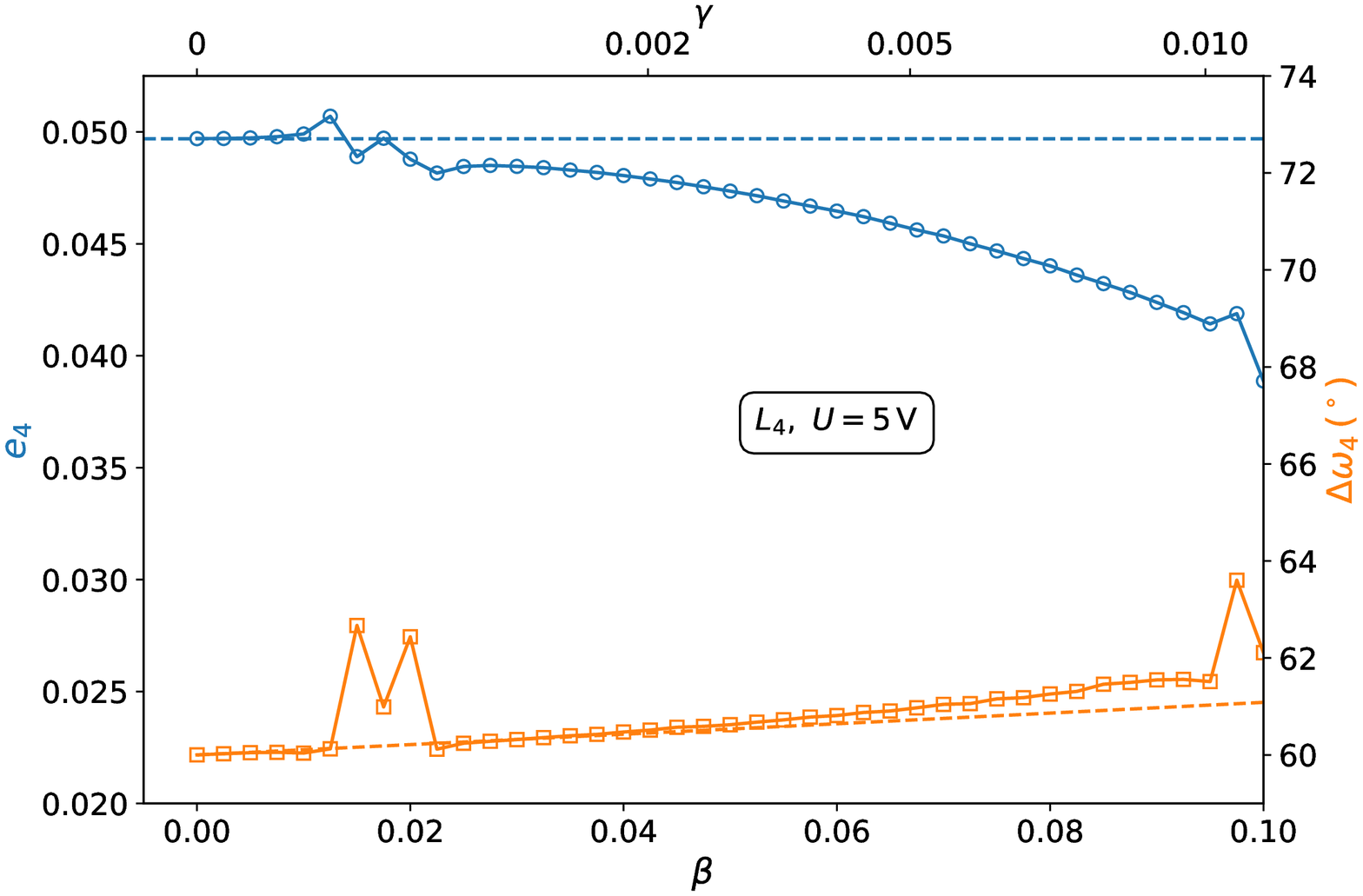}}
\resizebox{1.0\hsize}{!}{\includegraphics[width=\linewidth]{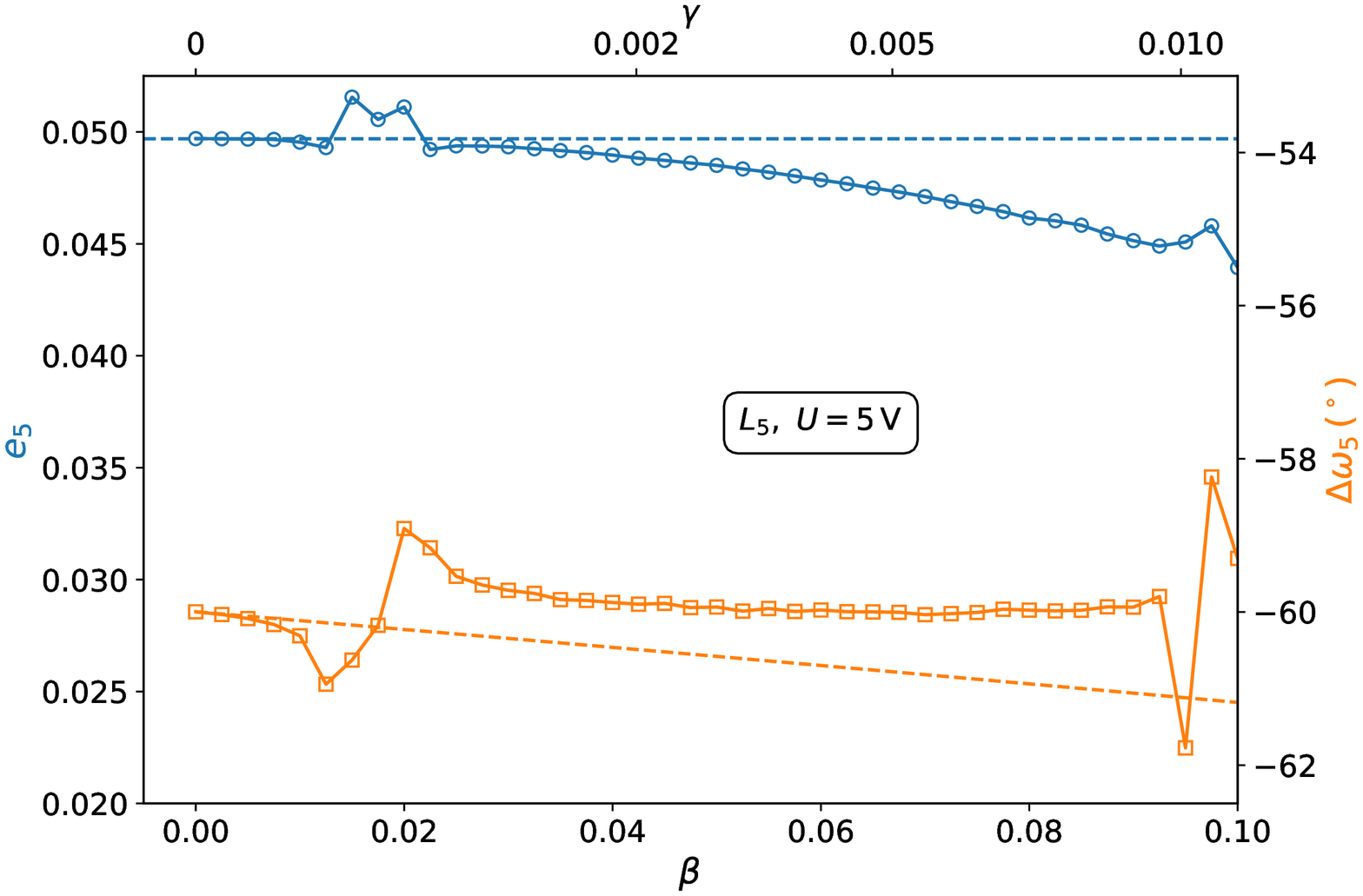}}
	\caption{Dependency of minimum libration amplitude
	in $e_k$ (blue) and $\Delta\omega_k$ (orange) in the vicinity of $L_k$
	(with $k=4$ on top and $k=5$ at the bottom) on parameter $\beta$ and
	$\gamma$ (the cases $\gamma=0$ are indicated by dashed lines) and dust
	grain surface potential $5$ Volts.}
\label{f:e-dw-Lk-u5}
\end{figure}

The results for minimum libration amplitude solutions concerning the variables
$\Delta\omega_k$ and $e_k$ are reported in Fig.~\ref{f:e-dw-Lk-u5}.
An important effect of charge, that is clearly visible, is the change of $e_k$ wrt.
to $\beta$ ($\gamma$) that cannot be seen in the uncharged problem, subject to
solar wind and the Poynting-Robertson effect alone. While for $\gamma=0$ we have
$e_k=e_J\simeq0.050$ we get $e_k\simeq0.040$ for $\beta=0.1$ (top of 
Fig.~\ref{f:e-dw-Lk-u5}) and a minimum in $e_k=0.045$ close to $L_5$ (at the 
bottom of the figure). The shift of $\Delta\omega_k$ for $\gamma>0$ also turns out
to be more prominent in comparison to the shifts in $\sigma_k$. As we can see
in Fig.~\ref{f:Delta-asig-kX} the maximum shift is about $2^\circ$ in $\Delta\omega_4$
(top) and about $2^\circ$ in $\Delta\omega_5$ (bottom).

\subsection{Resonance width and time of temporary capture}

\begin{figure} \centering
\resizebox{1.0\hsize}{!}{\includegraphics[width=\linewidth]{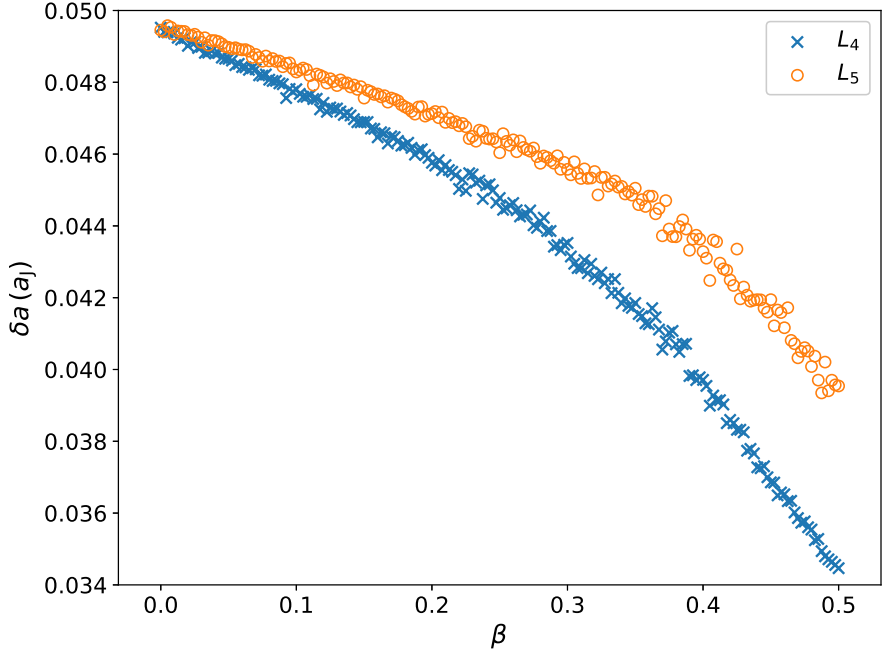}}
	\caption{Dependency of the extension of the regime of
	librational motion $\delta a$ (in untis of the semi-major axis of
	Jupiter $\delta_J$) on parameter $\beta$ in the uncharged problem.}
\label{f:lib-width}
\end{figure}

Once the initial conditions for the minimum libration amplitude solutions have been 
found we are interested in the extend of the librational regime of motions
around $L_4$ and in the vicinity of $L_5$ in dependency of parameters $\beta$
and $\gamma$. To estimate the `libration widths' we start by integrating 
\equ{e:eom} using the initial conditions from the previous study and increase
$a(0)=a^*+\delta a$ with $\sigma(0)=\sigma_k$ (with $k=4$ close to $L_4$ and
$k=5$ in case of $L_5$) until i) the maximum value of the resonant argument 
$\sigma(t)$ goes beyond $\sigma_5$ (thus $\sigma(t)>\sigma_5$) in case of
orbits originating from $L_4$ or ii) we find $\sigma(t)<\sigma_4$ for integrations 
starting around $L_5$. We report the size of the value $\delta a$ at which the crossing
of the librational regime takes place in Fig.~\ref{f:lib-width} for different 
values of the parameter $\beta$ and using $\gamma=0$. We clearly see that due to the 
additional perturbations $\delta a$ decreases with increasing value of $\beta$ with
a steeper slope in case of orbits originating from $L_4$ (blue crosses) in 
comparison to $L_5$ (orange circles). Thus, solar wind and Poynting-Robertson
effect also triggers the asymmetry between the extend of the two tadpole regimes of motion
around $L_4$ and $L_5$. \\

\begin{figure} \centering
\resizebox{1.0\hsize}{!}{\includegraphics[width=\linewidth]{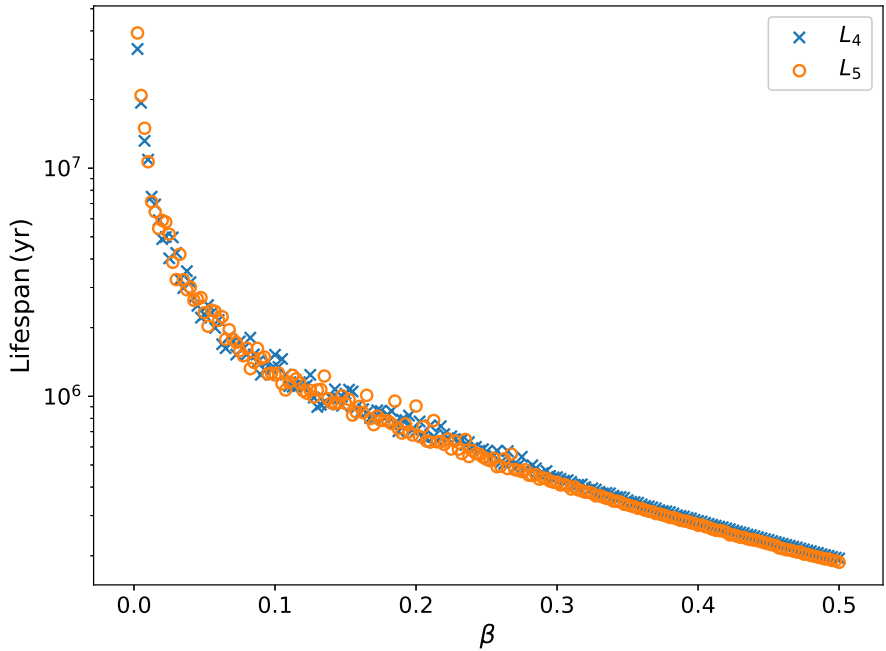}}
	\caption{Dependency of the time of temporary capture in
	$1:1$~MMR with planet Jupiter on parameter $\beta$ in the uncharged
	problem.}
\label{f:lifespan}
\end{figure}

The effect of charge on the width of the librational regime of motions is
clearly demonstrated in Fig.~\ref{f:dsigV}. However, the determination of the
maximum libration amplitude is not as straightforward as in the uncharged case.
The librational regime (indicated in blue) is tilted and irregular, and
moreover depends quite sensitively on the integration time. For this reason we
skip the study on the librational regime of motion in the charged problem and
investigate the time of temporary capture instead. For this reason, we
perform a series of numerical integrations for initial conditions starting at
the minimum libration amplitude location close to $L_4$ and $L_5$ and keep
track of the time of temporary capture in the vicinity of the $1:1$ MMR defined
by the time at which the dust grain leaves the regime of motion defined by
$|a(t)-a^*|>0.286$ AU.  The choice, that corresponds to about $0.055a_J$, is made
to cover the full tadpole regime of motions in the pure gravitational case
(see top of Fig.~\ref{f:a-sigb0b}). The results for the uncharged problem are reported in
Fig.~\ref{f:lifespan}.  Here, the capture time in mean motion resonance with
Jupiter starting close to $L_4$ is marked by blue crosses, the capture time for
orbits starting in the vicinity of $L_5$ is shown by orange circles.
Interestingly, the capture time, very close to the centers of the CRTBP, is 
more or less the same for both sets of initial conditions and changing $\beta$.

\begin{figure} \centering
\resizebox{1.0\hsize}{!}{\includegraphics[width=\linewidth]{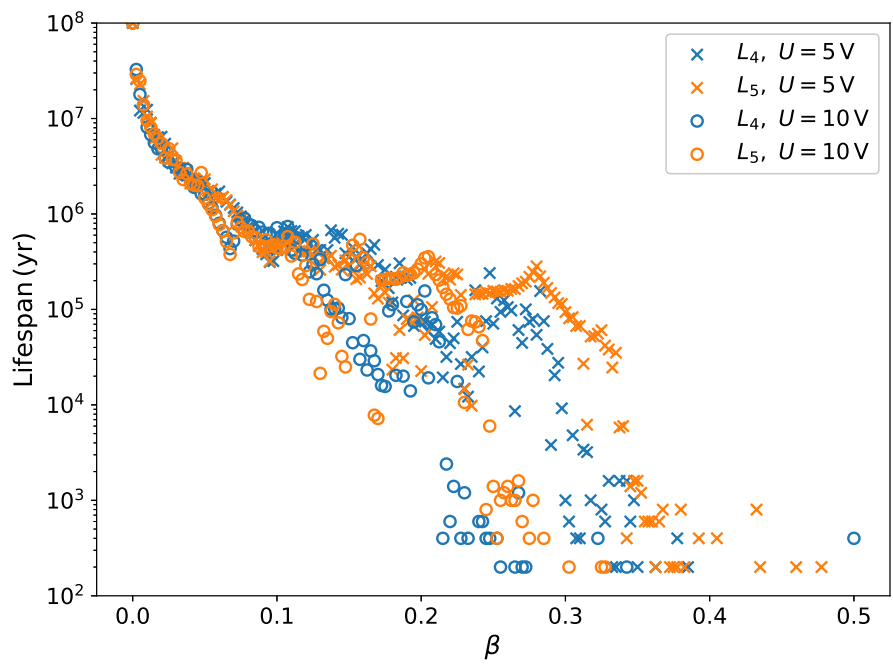}}
	\caption{Dependency of the time of temporary capture in
	$1:1$~MMR with planet Jupiter on parameter $\beta$ in the charged
	problem with surface potential of 5 and 10 Volts.}
\label{f:lifespan2}
\end{figure}

The results for the case $\gamma>0$, i.e. a dust surface charge potential of $5$ Volts
and $10$ Volts is shown in Fig.~\ref{f:lifespan2}. For small values of $\beta$ 
capture close to $L_4$ and $L_5$ takes place on comparable times with a slightly
large lifespan for $U=5V$ (crosses) compared to the case $U=10V$. Beyond 
$\beta\simeq0.1$ capture times behave more irregular, possibly due to the chaotic 
nature of the problem. However, we still see an increase in capture time
for the smaller value of dust grain surface potential $U$  (compare the location
of crosses and circles in Fig.~\ref{f:lifespan2}).

\section{A simplified model for charged co-orbital motion}
\label{s:ana}

In this section we develop a simplified mathematical model based on the 
circular restricted three-body problem and including Lorentz force, and
Poynting-Robertson effect.
It is valid very close to exact 1:1 mean motion resonance and is accurate 
for small values of eccentricity $e$ and orbital inclination $i$ of 
the charged dust particle only. The motivation of it is to understand
the role of the non-gravitational effects on $a$, $e$, and $i$ close
to equilibrium values. The model is obtained using perturbation 
theory and averaging method in the framework of Gauss' planetary equations 
of motion. Our focus is on the description of the mean evolution of 
$a$, $e$ and $i$ during time of temporary capture on secular time scales. 
Let $h=\sqrt{(1-e^2)}a^2n$, 
$n^2a^3=\mu$, $\theta=f+\omega$, with true anomaly $f$, and eccentric 
anomaly $E$. We start with the variational equations related to
semi-major axis, eccentricity, and inclination \citep[][]{2012icm..book.....F}:

\beqa{e:sm1}
\frac{da}{dt} &=&
\frac{2 h (F_T (e \cos (\theta )+1)+e F_R \sin (\theta ))}{\left(1-e^2\right) \mu }
\nonumber \\
\frac{de}{dt} &=&
\frac{h (F_T (\cos (\text{E})+\cos (\theta ))+F_R \sin (\theta ))}{\mu }
\nonumber \\
\frac{di}{dt} &=& 
\frac{F_N r \cos (\theta +\omega )}{h} \ .
\eeqa

\noindent where $F_R$, $F_T$, $F_N$ are the radial, tangential, and normal components of
the perturbed Kepler problem.  Here, the force components are defined in a coordinate
system with respect to the origin located at the position of the particle and the 
fundamental plane of reference coinciding with its orbital plane. The radial direction is
along the line connecting the sun and the particle, the tangential direction is along the 
velocity vector, and the normal component is orthogonal to the orbital plane. 
Denoting with index $j$ the components stemming
from i) a circular perturber ($j=1$), ii) from the Poynting-Robertson effect
($j=2$), and iii) the interaction of the charged particle with the
interplanetary magnetic field ($j=3$), these components can be split into the
form:

\beqa{e:sm2}
F_{R} &=& R_1 + R_2 + R_3 \nonumber \\
F_{T} &=& T_1 + T_2 + T_3 \nonumber \\
F_{N} &=& N_1 + N_2 + N_3 \ ,
\eeqa

\noindent and the field is related to the force functions $R_j$, $T_j$, $N_j$ by means of
\footnote{ See definition of rotation matrix ${\mathbf R_M}$ in \equ{e:sm4}.} 

\beqa{e:sm11}
R_j &=& \vec F_j\cdot {\mathbf R_M} \cdot 
\left(\cos\left(\theta\right), \sin\left(\theta\right), 0\right)^T \ , \nonumber \\
T_j &=& \vec F_j\cdot{\mathbf R_M} \cdot 
\left(-\sin\left(\theta\right), \cos\left(\theta\right), 0\right)^T \ , \nonumber \\
N_j &=& \vec F_j\cdot{\mathbf R_M} \cdot 
\left(0, 0, 1\right)^T \ ,
\eeqa

\noindent with $j=1,2,3$. Substitution of $R_i$, $T_i$, $N_i$ seperately into \equ{e:sm1}
results in the variations in orbital parameters $a$, $e$ and $i$ due to the
different perturbations. The functional form of each force component in
\equ{e:F} depends on $r$, $x$, $y$, $z$ and either on $v_x$, $v_y$, $v_z$
(Poynting-Robertson effect, Lorentz force) or on $r_1$, $x_1$, $y_1$, $z_1$
(the position of the circular perturber). First, we need to express these
quantities in terms of orbital elements. Assuming small values of $e$ and $i$
we start with the 2nd order expansions \citep[][]{myBook}:

\beqa{e:sm3}
\frac{r}{a} &=& 
1+\frac{1}{2}e^2-2e\sum_{\nu=1}^4
\frac{d J_\nu(\nu e)}{d e}\frac{\cos\left(\nu M\right)}{\nu^2} + O(e^3) \ ,
\nonumber \\
\cos\left(f\right) &=&
2\frac{1-e^2}{e}\sum_{\nu=1}^4
J_\nu(\nu e)\cos\left(\nu M\right) - e + O(e^3) \ , 
\nonumber \\
\sin\left(f\right) &=&
2\sqrt{1-e^2}\sum_{\nu=1}^4
\frac{d J_\nu(\nu e)}{de}\frac{\sin\left(\nu M\right)}{\nu} + O(e^3) \ ,
\nonumber \\
X &=&
2a\sum_{\nu=1}^4
\frac{d J_\nu(\nu e)}{de}\frac{\cos\left(\nu M\right)}{\nu^2}
-\frac{3ae}{2} + O(e^3) \ , 
\nonumber \\
Y &=&
2a\frac{\sqrt{1-e^2}}{e}\sum_{\nu=1}^4
J_\nu(\nu e)\frac{\sin\left(\nu M\right)}{\nu}
+O(e^3) \
\nonumber \\
\eeqa

\noindent (note that $Z=0$).  Here, $J_\nu=J_\nu(x)$, with integer index $\nu$, 
denote the Bessel functions of the 2nd kind. The transformation from the orbital frame
$(X, Y, Z)$ to the ecliptic frame $(x,y,z)$ is given by the rotation matrix
(see Fig.\ref{f:geo}):

\beq{e:sm4}
{\mathbf R_M} = {\mathbf R}_3(\Omega) \cdot {\mathbf R}_1(i) \cdot {\mathbf R}_3(\omega) \ ,
\eeq

\noindent  with rotation matrices ${\mathbf R_1}$, ${\mathbf R_3}$ and
using the transformation

\beq{e:sm5}
\left(x,y,z\right)^T = {\mathbf R_M}\cdot
\left(X,Y,Z\right)^T \ .
\eeq

\noindent The quantites $\dot r$, $\dot \vec r$ can be easily obtained 
from above relations by application of the operator $d/dt$ and taking into account
the dependency on time of mean anomaly $M=n t + M(0)$. To formulate $x_1$, $y_1$, $z_1$
in terms of Kepler elements, we make the assumption of a circular perturber 
($e_1=0$) moving within the ecliptic ($i_1=0$), and with vanishing perihel 
($\omega_1=0$) and longitude of the ascending node ($\Omega_1=0$). Denoting by $a_1$ 
the semi-major axis of the gravitational perturber, the above relations reduce to 
(note that $z_1=0$):

\beqa{e:sm6}
x_1 &=& a_1 \cos\left(M_1\right) \ , \nonumber \\
y_1 &=& a_1 \sin\left(M_1\right) \ ,
\eeqa

\noindent with $M_1=n_1 t$ and $n_1^2a_1^3=\mu$. For the expansion of the expression,

\beq{e:sm7}
\frac{\vec r_1 \cdot \vec r}{r_1^3}-\frac{1}{\Delta} - \frac{1}{r} \ ,
\eeq

\noindent that enters the potential in \equ{e:F} up to $O(e^3)$ we first approximate the 
term \equ{e:sm7} close to $r/r_1\simeq1$, and by making use of the identity:

\beq{e:sm8}
\cos\left(\psi\right) = \frac{\vec r \cdot \vec r_1}{r r_1} \ .
\eeq

\noindent Following the steps described in full detail in \citet{2015Icar..250..249L},
the terms that enter \equ{e:sm7} take the form

\beqa{e:sm9}
\frac{1}{\Delta}&=&\frac{1}{\sqrt{2}}\frac{1}{r_1}
\left(
\sum_{j=0}^\infty
\left(-1\right)^j\binom{-1/2}{j}\cos\left(\psi\right)^j
\sum_{n=0}^\infty
\binom{-1/2}{n}\epsilon^n
\right) \ , \nonumber \\
\frac{\vec r_1\cdot \vec r}{r_1^3} &=& \frac{r\cos\left(\psi\right)}{r_1^2} \ ,
\eeqa

\noindent with $\epsilon=r/r_1\left(1+A^{-1}r/r_1\right)$, and 
$A=2\left(1-\cos\left(\psi\right)\right)$. In the following, we make use of these 
expansions truncated at $(r/r_1)^2$ and $\cos^{12}\psi$. To obtain $\vec F_1$ we 
require the gradients of \equ{e:sm9} that become:

\beqno
\frac{dr}{dx} = \frac{x}{r} \ , 
\frac{dr}{dy} = \frac{y}{r} \ ,
\frac{dr}{dz} = \frac{z}{r} \ ,
\eeqno

\noindent as well as

\beqano
\frac{d\cos\left(\psi\right)}{dx} = \frac{x_1}{r r_1} \ ,
\frac{d\cos\left(\psi\right)}{dy} = \frac{y_1}{r r_1} \ ,
\frac{d\cos\left(\psi\right)}{dz} = \frac{z_1}{r r_1} \ .
\eeqano

\noindent Using \equ{e:sm3} together with above expressions the vector field
$\vec F_1$ is completely determined by the gradient of the potential

\beq{e:sm10}
\Phi_1 = - \mu_1\left(\frac{\vec r_1.\vec r}{r_1^3}-\frac{1}{\Delta} - \frac{1}{r} \right) \ ,
\eeq

\noindent We insert \equ{e:sm2}, \equ{e:sm11} in \equ{e:sm1} using \equ{e:sm3}
- \equ{e:sm10}, and $\vec F_1=-\nabla \Phi_1$, and only retain trigonometric
terms in the expansions of \equ{e:sm1} that are of the form:

\beqno
k(M+\omega+\Omega-M_1)+l\omega+m\Omega \ ,
\eeqno

\noindent with $(k, l, m)\in\mathbb{Z}^3$. The resulting vector field only
contains resonant terms that are trigonometric in resonant argument
$\sigma=\lambda-\lambda_1$, with $\lambda=M+\omega+\Omega$, and without
explicit dependence on the orbital longitude of the perturber $\lambda_1=M_1$.
Let supscript ${}^{(1)}$ label the orbital variation due to the perturber. To
2nd order in $r/r_1$, 12th order in $\cos(\psi)$ and 2nd order in $e$ we obtain
the system:

\beqa{e:sm12}
\frac{da^{(1)}}{dt}&=&
\sum_{k,l,m}c_{k,l,m}^{(a)}(e)\cos\left(k i\right)\sin\left(l\sigma+m\omega\right) \ , 
\nonumber \\
\frac{de^{(1)}}{dt}&=&
\sum_{k,l,m}c_{k,l,m}^{(e)}(e)\cos\left(k i\right)\sin\left(l\sigma+m\omega\right) \ ,
\nonumber \\
\frac{di^{(1)}}{dt}&=&
\sum_{k,l,m}c_{k,l,m}^{(i)}(e)\sin\left(k i\right)\sin\left(l\sigma+m\omega\right) \ ,
\eeqa

\noindent with $k,l=1,\dots,12$, $m=-4,-2,0,2,4$, and $c_{k,l,m}^{(a,e,i)}$
polynomial in eccentricity $e$. We notice that the vector field does not explicitely depend on
ascending node longitude $\Omega$. From standard theory of the circular
restricted three-body problem we know \citep[][]{myBook} that the reference
solution for the stable equilibria $L_4$, $L_5$ is given by $a_*=a_J$,
$e_*=e_J=0$, $i_*=i_J=0$, $\omega_*=\omega_J=0$, $\Omega_*=\Omega_J=0$, and
$M_*=M_J\pm60^o$.  The solution defined by the pure gravitational problem from
\equ{e:sm12} is determined by the condition:

\beq{e:sm13}
\frac{da^{(1)}}{dt} = \frac{de^{(1)}}{dt} = \frac{di^{(1)}}{dt} = 0 \ .
\eeq

\noindent To solve this system we substitute $M_*=\pm60^o$, and
$\omega_*=\Omega_*=0$ and solve for the remaining $a_*^{(1)}$, $e_*^{(1)}$,
$i_*^{(1)}$ using a numerical scheme (Newton method). A comparison with the
reference solution provides an estimate of the error in the approximation of
the exact problem. 

\subsection{The role of radiative effects on shift in $a$}

The shift due to parameter $\beta$ that enters $\vec F_0$ in 
\equ{e:eom} has been found to follow $a=a_J\left(1-\beta\right)^{1/3}$, the
shift in $\sigma$ is given in the upper plot of Fig.~\ref{f:sigdw-45}, obtained
numerically, and on the basis of a simplified formula in synodic coordinates
\citep[][]{2021A&A...645A..63Z}.

To estimate the role of the  combined PR and solar wind effect
on the Kepler elements we make use of Eq.(14) in \citet{2015Icar..250..249L}, where
the secular effect on semi-major axis $a$ and eccentricity $e$ is simply given by:

\beqa{e:sm13}
\frac{da^{(2)}}{dt} &=& -\frac{a(1+3e^2)\mu\beta n}{c(1-e^2)^{3/2}} \ , \nonumber \\
\frac{de^{(2)}}{dt} &=& -\frac{\mu\beta n}{\sqrt{a}e c}
\left(\frac{(3+2e)e}{2\sqrt{1-e^2}}\right) \ ,
\eeqa

\noindent and $di^{(2)}/dt = 0$ (supscript ${}^{(2)}$ indicates again the link of
\equ{e:sm13} with $\vec F_2$). We notice that radiative effects do not affect the orbital
planes, and the signs that enter in front of \equ{e:sm13} indicate
that the orbits of the dust particles are shrinking in $a$ and circularizing in 
$e$ with time. Taking $n=\sqrt{\mu/a^{3/2}}$ and then
$a=a_J\left(1-\beta\right)^{1/3}$  a Taylor series expansion
with respect to $\beta$ gives to zeroth order in $e$ the estimate:

\beqno
\frac{da^{(2)}}{dt} \simeq -\frac{\mu^{3/2}}{\sqrt{a_J}c}\left(
\beta+\frac{\beta^2}{6}\dots\right) \lesssim
1.9\times10^{-5}\beta \ ,
\eeqno

\noindent and $de^{(2)}/dt \simeq 0 $, where we substituted for the parameters
$\mu$, $a_J$, $c$ to obtain the inequality. The magnitudes being small,
the shift of the equilibrium values in $a$ and $e$ can be neglected
in comparison to the shift $a_J(1-\beta)^{1/3}$.
However, the effect is sufficient to render the equilibria unstable, 
and are therefore responsible for the phenomenon of temporary 
capture close to $L_4$ and $L_5$, respectively
\citep[see, e.g.][]{1994Icar..112..465M,
2015Icar..250..249L}.

\subsection{The role of the interplanetary magnetic field on $i$}

To model the mean effect on $a$, $e$, $i$ that is stemming from 
the interaction of the charged dust particle with the interplanetary magnetic field, 
i.e. $\vec F_3$ in \equ{e:eom}, we proceed as follows. Assuming a standard Parker 
spiral model \citep[][]{2019AnGeo..37..299L} of the mean magnetic field we make use of the 
expressions developed in \citet{2019CeMDA.131...49L}, i.e. Eq. (24):

\beq{e:sm14}
\frac{di^{(3)}}{dt} =
-\alpha\frac{q}{m}\frac{B_0}{2}\left(\frac{r_0}{a}\right)^2\bigg\{
\left[1 - \cos\left(i\right)z_0 \frac{\Omega_s}{n}\right]\times
\bigg(x_0\cos\left(\Omega\right)+y_0\sin\left(\Omega\right)\bigg)\bigg\} \ .
\eeq

\noindent Here, $x_0=\sin(\Omega_0)\sin(i_0)$, $y_0=-\sin(i_0)\cos(\Omega_0)$,
$z_0=\cos(i_0)$, that locate the magnetic dipole axis of the rotating sun in
the inertial reference frame, see Fig.~\ref{f:geo}. We notice that \equ{e:sm14}
is valid for small $x_0$, $y_0$, and $1-z_0$ and vanishing eccentricty only.
Moreover, we essentially neglect the influence of the radial and tangent
components $F_R$, $F_T$ in \equ{e:sm1}, that have been shown to vanish over one
revolution period of the dust particle \citep[][]{2019CeMDA.131...49L}.
However, we stress that the modification of the standard Parker spiral model to
include a magnetic field normal component $B_N$ will result in secular
evolution of $da/dt$, $de/dt$, and $di/dt$ \citep[see, Eq. (14)
in][]{2016ApJ...828...10L}.  Since we assume $B_N=0$ throughout the paper we
may use \equ{e:sm14} to investigate the role of Lorentz force on the orbital
evolution of inclination of the dust particles. 

\begin{figure} \centering
\resizebox{1.0\hsize}{!}{\includegraphics[width=0.95\linewidth]{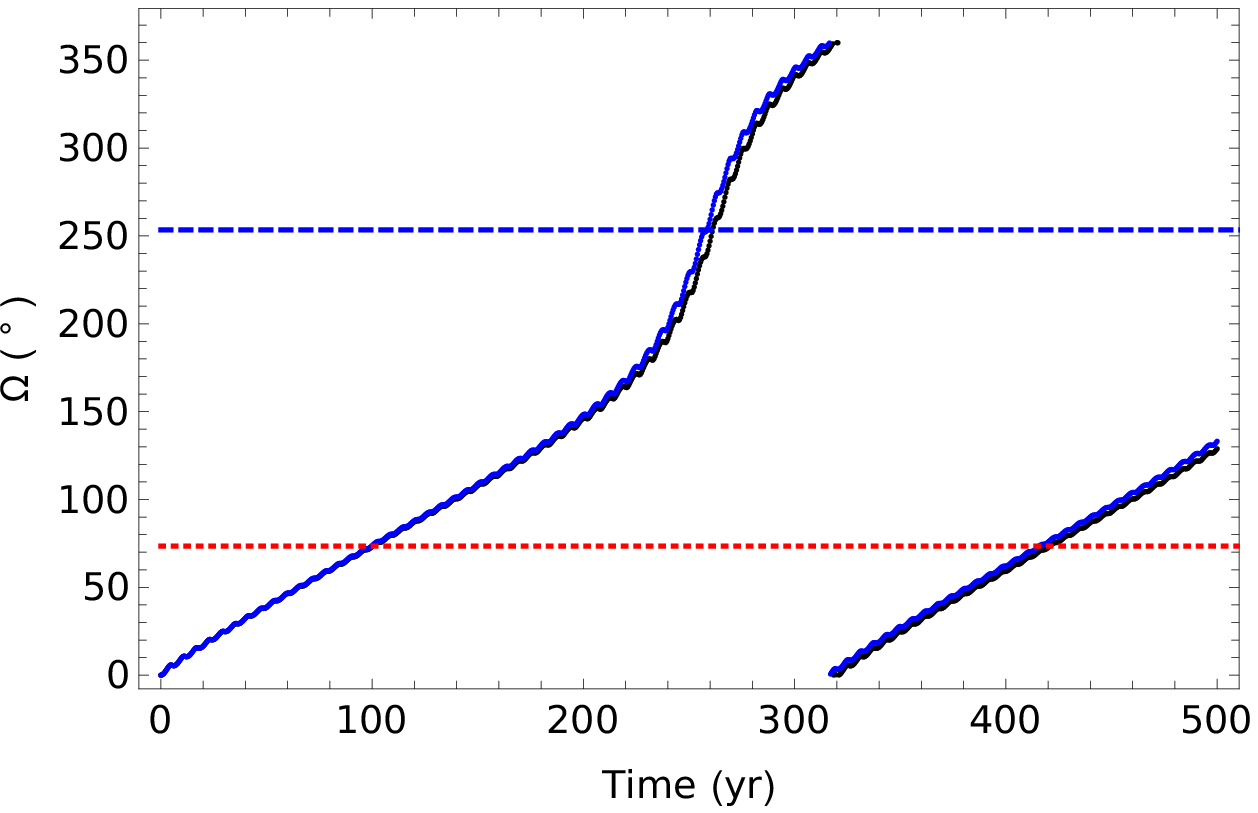}}
\resizebox{1.0\hsize}{!}{\includegraphics[width=0.95\linewidth]{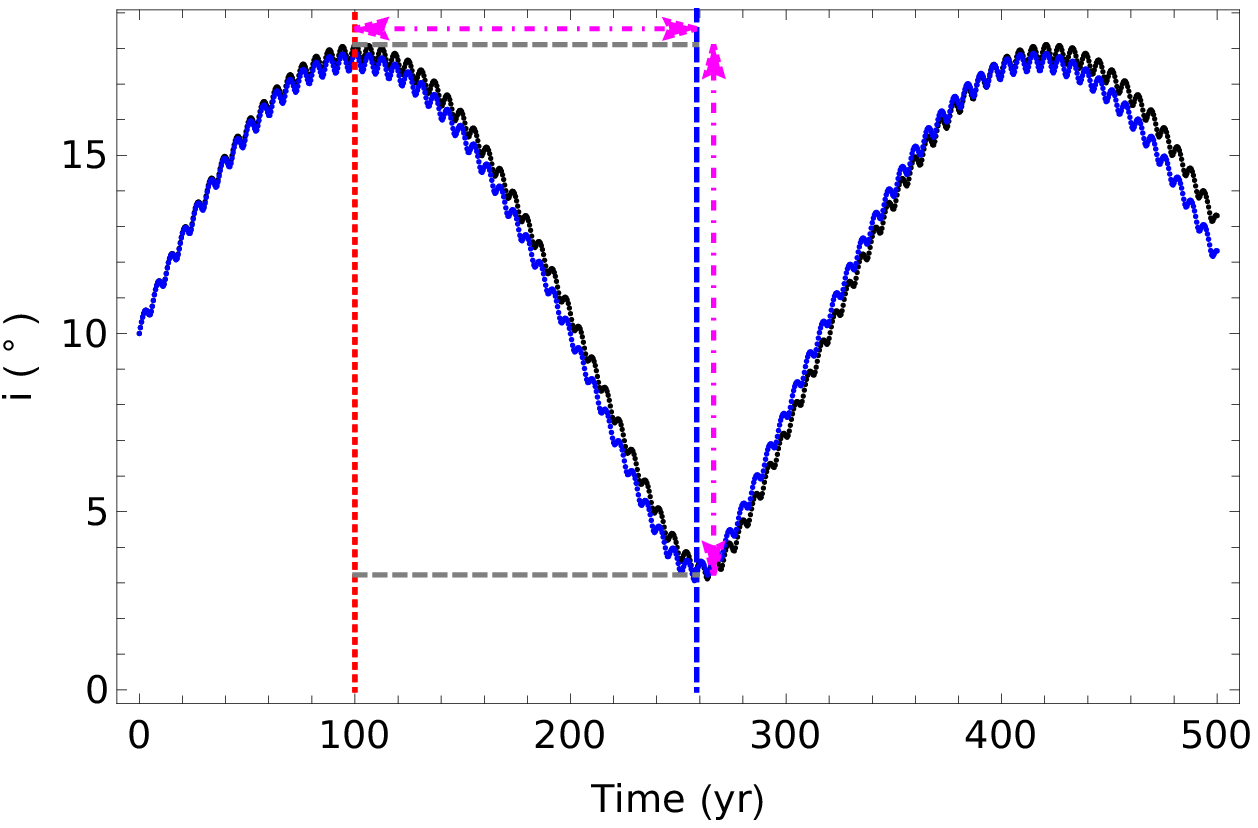}}
	\caption{Timeseries for ascending node longitude $\Omega$ (top) and
	inclination (bottom) for a dust particle of radius $R=2.05\mu m$ with $4.43V$ surface
	charge, starting at $a(0)=a_J$, $e=0.01$, $i(0)=10^o$. Red-dotted lines mark
	$\Omega=\Omega_0=73.5^o$ where $i=i_{max}$, blue-dashed lines 
	$\Omega=\Omega_0+180=253.5^o$, 
	where $i=i_{min}$. Arrows (dot-dashed magenta) indicate $i_{max}-i_{min}\simeq15^o$,
	and the timespan between maxima and minima (about $160$ years).}
\label{f:nod-inc}
\end{figure}

From numerical studies in Sec.~\ref{s:stud} we already found that the inclusion
of the interplanetary magnetic field on the dynamics triggers periodic
variations of the orbital planes of the charged dust particles as shown on the
lower left in Fig.~\ref{f:XH}. Following the approach developed in
\cite{2019CeMDA.131...49L} we estimate the net effect on the orbital
inclination as follows. From the condition
$x_0\cos\left(\Omega\right)+y_0\sin\left(\Omega\right)=0$
we find $\Omega=\Omega_0$ and as a consequence $di/dt=0$ in \equ{e:sm14}.
The turning points in $i$ should therefore take place whenever the line of nodes
related to ascending node longitudes coincide with the line of nodes formed 
between the equatorial and ecliptic planes located at angular distance 
$\Omega=\Omega_0=73.5^o$. The result is visualized in Fig.\ref{f:nod-inc} for
a test particle starting with $i(0)=10^o$. At the beginning the inclination of 
the charged dust particle increases. After a period of time of about
$100y$ the ascending node of the dust particle crosses $\Omega=\Omega_0=73.5^o$,
where we have $di/dt=0$ and the maximum excursion in $i=i_{max}\simeq18^o$. The
effect of Lorentz force reverses, and inclination decreases until the minimum
$i_{min}\simeq3^o$ is reached at time $t\simeq260y$, where ascending node 
longitude passes $\Omega=\Omega_0+180^o$, and inclination starts to rise again.
The simple analysis on the basis of \equ{e:sm14} provides libration amplitudes 
in inclination $i$ of about $15^o$ with libration period of about $320y$ which
is consistent with numerical simulations.

\section{Summary \& Conclusions}
\label{s:sum}

In this work we study the dynamics of charged dust close to the Lagrangian
points $L_4$ and $L_5$ and subject to solar wind, Poynting-Robertson (PR)
effect, and the interplanetary magnetic field, with special focus on the
Lorentz force term.  We provide the shift and extent of the tadpole regime of
motions in dependency on the system parameters, i.e. the charge-to-mass ratio
$\gamma$ of the dust grains.  We quantify the asymmetry between the location
and size of librational kind of motions between $L_4$ and $L_5$ which is mainly
due to radiative effects and the Lorentz force term.  The shift in resonant
argument $\sigma$ due to the solar wind and the PR-effect from the pure
gravitational solution $\sigma=\pm60^o$ is larger with respect to $L_4$ in
comparison with $L_5$. For small charge-to-mass ratios the shift due to Lorentz
force is small, but may increase to several degrees with decreasing radius $R$
of the charged dust grain. A similar behaviour can be found in the angular
spearation $\Delta \omega=\omega-\omega_J$ from the value $\pm60^o$ of the pure
gravitational, elliptic problem. The displacement in semi-major axis $a$ from
$a_J$ of Jupiter is dominated by solar radiation pressure, following
$a\simeq(1-\beta)^{(1/3)}a_J$, with increasing ratio $\beta$ between solar
radiation over pure gravitational attraction. However, it is found that Lorentz
force may contribute to this displacement with increasing values of $\gamma$.
One important finding of our study is the role of $\gamma$ on $\Delta e =
e-e_J$ that vanishes at the Lagrange points of the pure gravitational problem.
While radiation pressure, solar wind, and the PR-effect leads to marginal
variations in $\Delta e$, Lorentz force may trigger large deviations from zero
at the minimum libration amplitude solutions. Another important phenomenon that
can only be explained by the Lorentz force, i.e. the interaction of the charged
dust grains with the interplanetary magnetic field, are periodic variations on
secular time scales in inclination $i$ and ascending node longitude $\Omega$.
Using a simplified model on the basis of Gauss averaged equations of motion,
conditions for maxima and minima of these excursions in inclinations can be
found with libration amplitudes up to several degrees. We also observe
amplitude enhancements in the orbital evolution of the resonant argument
$\sigma$ whenever the periods of ascending node longitude $\Omega$ and $\sigma$
coincide. Since the period in $\Omega$ depends on the actual charge-to-mass
ratio $\gamma$ the phenomenon occurs at specific values of $\gamma$ only.
Last, but not least we perform a series of simulations to estimate the time of
temporary capture of charged dust close to the 1:1 mean motion resonance with
planet Jupiter, and find a decrease in capture time for charged dust in
comparison to neutral one. We notice that our model of the interplanetary
magnetic field is very simple, i.e. it does not include time dependent effects
stemming from the activity of the Sun. Our results are therefore not suitable
to interpret observations.  However, even this simple model already shows the
important role of the solar wind, together with the interplanetary magnetic
field on resonant kind of motions of charged dust in planetary systems. More
realistic models of the interplanetary magnetic field, and time dependent
effects will be subject to future studies in the field.

\vskip.1in

{\bf Acknowledgements} This work is funded by the Austrian Science Fund
(FWF) within the project P-30542 entitled 'Stability of charge and orbit of 
cosmic dust particles'. CL acknowledges the support of 
EU H2020 MSCA ETN Stardust-Reloaded Grant Agreement 813644,
MIUR Excellence Department Project awarded to the Department of Mathematics, 
University of Rome Tor Vergata, CUP E83C18000100006, MIUR-PRIN 20178CJA2B 
'New Frontiers of Celestial Mechanics: theory and Applications', and GNFM/INdAM.
LZ acknowledges the support of China Scholarship Council (No. 201906190106),
National Natural Science Foundation of China (NSFC, Grants No. 11473016, No. 11933001),
and National Key R\&D Program of China (2019YFA0706601).\\

The authors declare that they have no conflict of interest.

\bibliographystyle{plainnat}
\bibliography{biblio}

\end{document}